\documentclass[acmsmall,screen]{acmart}
\usepackage{amsfonts}
\usepackage{wrapfig}

\usepackage{tcolorbox}
\tcbuselibrary{listings,skins,breakable}

\usepackage{listings}

\usepackage{tikz}
\usetikzlibrary{shapes.geometric,arrows.meta,positioning,calc,backgrounds,fit}
\usepackage{amsmath}
\usepackage{stmaryrd}  
\usepackage{mathtools}

\usepackage{filecontents}

\usepackage{graphicx}
\usepackage{subcaption}

\usepackage{caption} 
\usepackage{lipsum}

\usepackage{tabularx}
\usepackage{booktabs}
\usepackage{hyperref}
\usepackage{amssymb}
\usepackage{enumitem}
\usepackage{comment}
\usepackage{algorithm}
\usepackage[noend]{algpseudocode}
\usepackage{microtype}
\usepackage{xspace}
\usepackage{makecell}
\usepackage{listings}
\usepackage{xcolor}
\usepackage{minted}

\usepackage[table]{xcolor}

\DeclarePairedDelimiter{\ceil}{\lceil}{\rceil}
 
\definecolor{codegreen}{rgb}{0.4,0.6,0.2}
\definecolor{codepurple}{rgb}{0.6,0.2,0.6}
\definecolor{codeblue}{rgb}{0.1,0.3,0.6}
\definecolor{codered}{rgb}{0.7,0.1,0.1}
\definecolor{codeteal}{rgb}{0.0,0.5,0.5}
\definecolor{codegray}{rgb}{0.5,0.5,0.5}
\definecolor{codebg}{rgb}{0.97,0.97,0.97}

\lstdefinestyle{pythonstyle}{
     language=Python,
     backgroundcolor=\color{codebg},
     basicstyle=\ttfamily\tiny,
     keywordstyle=\color{codepurple}\bfseries,
     stringstyle=\color{codered},
     commentstyle=\color{codegreen},
     numberstyle=\tiny\color{codegray},
     breaklines=true,
     frame=single,
     framerule=0.5pt,
     xleftmargin=2pt,
     xrightmargin=2pt,
     aboveskip=3pt,
     belowskip=3pt,
     showstringspaces=false,
     morekeywords={def,return,True,False},
     emph={semantics,precondition,SymTensor,OpSema,SymAxis,             tile_constraint},
     emphstyle=\color{codeteal},
     emph={[2]nc_matmul,mul,sum},
     emphstyle={[2]\color{codeblue}},
}

\newcommand{\hmc}[1]{%
  \begingroup
  \edef\val{#1}%
  \ifdim\val pt<1.0pt
    \cellcolor{red!25}\val
  \else\ifdim\val pt<1.1pt
    \cellcolor{blue!15}\val
  \else\ifdim\val pt<1.2pt
    \cellcolor{blue!30}\val
  \else
    \cellcolor{blue!50}\val
  \fi\fi\fi
  \endgroup
}

\newcommand{\hm}[1]{%
  \begingroup
  \edef\val{#1}%
  \ifdim\val pt<1.5pt
    \cellcolor{blue!15}\val
  \else\ifdim\val pt<2.5001pt
    \cellcolor{blue!30}\val
  \else
    \cellcolor{blue!50}\val
  \fi\fi
  \endgroup
}

\newcommand{\pname}{\textsc{Axon}\xspace}
\begin{document}

\title{\pname{}: A Synthesizing Superoptimizer for Tensor Programs}
\author{Akash Kothari}
\affiliation{University of Illinois at Urbana-Champaign\country{USA}
}
\author{Shaowei Zhu}
\affiliation{Amazon Web Services\country{USA}
}
\author{Daniel Kroening}
\affiliation{Amazon Web Services\country{USA}
}
%\email{shaowz@amazon.com}
\author{Chungha Sung}
\affiliation{Amazon Web Services\country{USA}
}
%\email{chunghs@amazon.com}
%\affiliation{%
%  \institution{Institute for Clarity in Documentation}
%  \country{USA}
%}

%\affiliation{%
%  \institution{Inria Paris-Rocquencourt}


\begin{abstract}
Writing high-performance kernels for AI accelerators requires deep expertise in tiling, instruction selection, data layout, and operator fusion---placing a significant burden on programmers.

In this paper, we focus on tile-based AI accelerator programs and present \pname{}, a synthesizing superoptimizer for tensor programs: it uses program synthesis to automatically generate target instructions from semantics specifications, and explores semantically equivalent program variants to select the best-performing kernel empirically.
\pname{} discovers algebraic transformations by propagating operators through computation graphs and uses SMT over unbounded tensors to guarantee that all transformations preserve semantics---without requiring hand-crafted rewrite rules.
It then lowers tensor operations to target ISA instructions, explores tiling configurations constrained by hardware descriptions, and fuses operators and instructions to minimize memory traffic.

We evaluate \pname{} on Amazon's Trainium across 20 benchmarks spanning individual tensor operators and multi-operator LLM kernels, including Group Query Attention and Gated MLP.
Compared to Amazon's Neuron compiler, \pname{} achieves speedups of up to 3.7x on individual operators and up to 19x on multi-operator kernels.
\pname{} matches or outperforms hand-optimized NKI kernels by up to 1.35x, and achieves geomean speedups of 2--10\% over Mirage, a state-of-the-art search-based compiler, on supported benchmarks---while synthesizing kernels for 16 additional benchmarks that Mirage cannot handle.
\end{abstract}

\maketitle

\section{Introduction}
\label{sec:intro}

Performance of modern AI accelerators depends critically on the quality of kernels---low-level programs that describe how hardware resources should be orchestrated to execute a workload.
Even though tile-based kernel languages (e.g., NKI for Amazon's Trainium~\cite{aws_neuron_nki_index}, Pallas for Google's TPUs~\cite{pallas}, and Triton~\cite{tillet2019triton} for GPUs) provide easier pathways to good performance by abstracting away many low-level details, it remains challenging to write kernels that achieve high utilization of hardware.
Kernel programmers still need to understand hardware details such as memory hierarchies and specialized compute and data movement engines to reason about tiling, data layout, operator fusion, and engine selection---tasks that require deep hardware expertise and are out of reach for most ML engineers.

\begin{wrapfigure}{r}{0.50\textwidth}
\centering
\vspace{-20pt}
\includegraphics[width=0.52\textwidth]{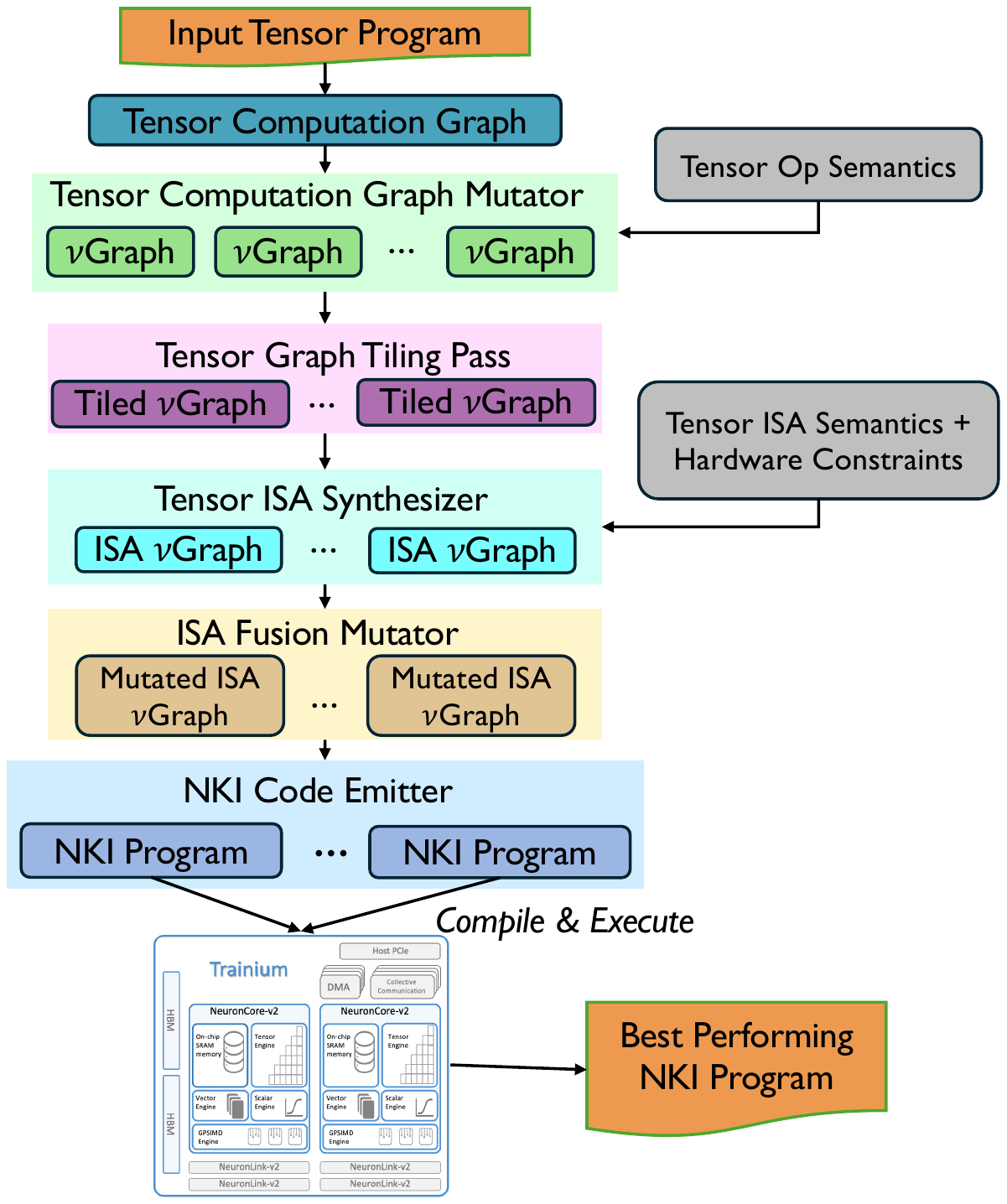}
\vspace{-.5in}
\caption{Overview of \pname{}'s workflow.}
%\vspace{-8pt}
\label{fig:axon-workflow}
\end{wrapfigure}

We present \pname{}, a synthesizing superoptimizer for tile-based AI accelerator programs: given tensor programs written in a Numpy-like interface~\cite{harris2020array}, it uses program synthesis to automatically generate target instructions from semantics specifications, and explores all semantically equivalent program variants to select the best-performing kernel empirically (Figure~\ref{fig:axon-workflow}).

%it uses program synthesis to automatically generate target instructions from semantics specifications, and explores all semantically equivalent program variants to select the best-performing kernel empirically.
%
%\pname{} enables programmers to write tensor programs using a NumPy-like interface~\cite{harris2020array} and automatically generates optimized kernels for the target hardware.
%
%Figure~\ref{fig:axon-workflow} shows the overall workflow.

\pname{} discovers algebraic transformations by \textit{propagating} operators through computation graphs, reordering operations to expose parallelism and fusion opportunities.
To guarantee that all transformations preserve semantics, \pname{} decomposes tensor operations into granular computations and uses an SMT solver to check equivalence over unbounded tensors---requiring no hand-crafted rewrite rules or operator properties.
For example, in an RMSNorm+MatMul kernel, \pname{} automatically discovers that an element-wise multiply can be reordered past the matrix multiplication, enabling the normalization and matmul to execute on different compute engines in parallel (Section~\ref{sec:motivation}).

\pname{} then lowers each tensor operation to target ISA instructions via sketch-driven program synthesis, using semantics specifications that describe each instruction's computation and hardware constraints.
It explores tiling configurations using symbolic parameters constrained by the hardware description, and fuses operators and instructions to minimize memory traffic.
All valid variants---across algebraic transformations, instruction selections, tiling configurations, and fusion choices---are maintained simultaneously in a $\nu$Graph (``nu-graph''), a persistent internal representation that tracks all semantically equivalent program variants through each stage of the synthesis pipeline.
The best-performing kernel is selected empirically when input shapes are known.

Existing approaches to tensor program optimization either require hand-crafted rewrite rules for algebraic transformations~\cite{jia2019taso, tensat, wang2021pet}, rely on hand-written templates for code generation~\cite{wu2025mirage}, or target only CPUs and GPUs~\cite{chen2018tvm, zheng2020ansor, ragan2013halide}.
\pname{} derives all transformations and instruction selections from semantics specifications, enabling it to automatically generate optimized kernels for tile-based AI accelerators with provable correctness guarantees.

\par \textbf{Contributions.} This paper makes the following contributions:
\begin{itemize}
    \item A \textbf{provably correct methodology for algebraic transformations} through operator propagation, which automatically discovers semantically equivalent computation graphs by decomposing tensor operations into granular computations and using SMT over unbounded tensors. It checks that reordering adjacent operators preserves semantics---requiring no hand-crafted rewrite rules or operator properties.

    \item A \textbf{synthesis-based pipeline} for tile-based AI accelerator programs that automatically lowers high-level tensor operations to target ISA instructions, explores tiling configurations constrained by hardware descriptions, and performs operator and instruction fusion. All semantically equivalent variants are maintained simultaneously and the best-performing kernel is selected empirically.

    \item An \textbf{evaluation on Amazon's Trainium} across 20 benchmarks spanning individual operators and multi-operator LLM kernels (including Group Query Attention and Gated MLP), demonstrating speedups of up to 3.7x on individual operators and up to 19x on multi-operator kernels over Amazon's Neuron compiler, up to 1.35x over hand-optimized NKI kernels, and geomean speedups of 2--10\% over Mirage~\cite{wu2025mirage} on all benchmarks it supports---while synthesizing 16 additional kernels that Mirage cannot.
\end{itemize}

\section{Motivation}%: RMSNorm + Matmul Kernel}
\label{sec:motivation}

\begin{wrapfigure}{r}{0.5\textwidth}
\centering
\vspace{-.2in}
\includegraphics[width=0.5\textwidth]{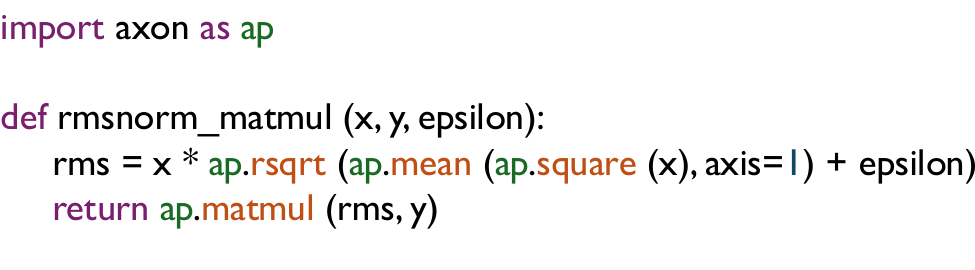}
\vspace{-.3in}
\caption{An \pname{} program for RMSNorm+MatMul using NumPy-like interfaces.
}
%\vspace{-10pt}
\label{fig:ap-example}
\end{wrapfigure}

We illustrate \pname{}'s approach using a kernel composed of RMSNorm~\cite{rmsnorm} followed by a matrix multiplication---a computation pattern that appears multiple times per transformer layer in modern LLMs such as LLaMA-3~\cite{grattafiori2024llama}, where RMSNorm precedes each linear projection (QKV, output, and FFN layers).
Figure~\ref{fig:ap-example} shows how a programmer expresses this kernel in NumPy-like~\cite{harris2020array} interface; the rest of this section walks through how \pname{} automatically optimizes it.

RMSNorm normalizes an input tensor $X$ by its root mean square (RMS) along a particular dimension (typically the last dimension of~$X$):
\begin{equation}
\text{RMSNorm}(X) = X \odot \text{RMS}(X) = X \odot \frac{1}{\sqrt{\frac{1}{N}\sum_{i} X_i^2 + \epsilon}}
\end{equation}
where $\epsilon > 0$ is a small constant for numerical stability.
The output is then matrix-multiplied with a weight tensor $W$ to produce the final result $Z$.

\begin{wrapfigure}{r}{0.5\textwidth}
\centering
\vspace{-13pt}
\raisebox{0pt}[\height][0pt]{\begin{tikzpicture}[
    scale=0.45,
    transform shape,
    input/.style={rectangle, draw=purple!70!black, fill=purple!30, minimum width=0.8cm, minimum height=0.7cm, font=\Large},
    output/.style={rectangle, draw=olive!70!black, fill=yellow!40, minimum width=0.8cm, minimum height=0.7cm, font=\Large},
    rmsop/.style={ellipse, draw=cyan!60!black, fill=cyan!40, minimum width=1.6cm, minimum height=0.9cm, font=\Large},
    mulop/.style={ellipse, draw=green!60!black, fill=green!35, minimum width=1.6cm, minimum height=0.9cm, font=\Large},
    matmulop/.style={ellipse, draw=red!40!orange, fill=red!25!orange!30, minimum width=1.6cm, minimum height=0.9cm, font=\Large},
    ghostop/.style={ellipse, draw=gray!60, dashed, fill=darkgray!20, minimum width=1.6cm, minimum height=0.9cm, font=\Large, text=black!70},
    arrow/.style={->, >=stealth},
    redarrow/.style={->, >=stealth, red!70, dashed},
    dashbox/.style={draw, dashed, rounded corners=5pt, inner sep=6pt}
]

% Part (a)
\begin{scope}[local bounding box=parta]
    \fill[gray!15, rounded corners] (-0.7,1.4) rectangle (14.8,-3);

    \node[input] (eps) at (0, 0.7) {$\epsilon$};
    \node[input] (x1) at (0, -0.3) {$X$};
    \node[input] (y1) at (0, -2.3) {$W$};

    \node[rmsop] (sq) at (2, 0) {square};
    \node[rmsop] (mean) at (4, 0) {mean};
    \node[rmsop] (add) at (6, 0) {add};
    \node[rmsop] (rsqrt) at (8, 0) {rsqrt};

    \node[mulop] (mul) at (10, -1.2) {multiply};
    \node[matmulop] (mm) at (12, -2.4) {matmul};
    \node[output] (z1) at (14, -2.4) {$Z$};

    \draw[dashbox, black!60] (0.7, 1.1) rectangle (11.5, -1.8);
    \node[font=\Large\itshape, anchor=north east] at (11.5, 1.05) {RMSNorm};

    \draw[arrow] (x1.east) to[out=0, in=180] (sq.west);
    \draw[arrow] (sq) -- (mean);
    \draw[arrow] (mean) -- (add);
    \draw[arrow] (eps.east) to[out=10, in=160] (add.north west);
    \draw[arrow] (add) -- (rsqrt);
    \draw[arrow] (rsqrt.east) to[out=0, in=90] (mul.north);
    \draw[arrow] (x1.east) to[out=-20, in=180] (mul.west);
    \draw[arrow] (mul) -- (mm);
    \draw[arrow] (y1.east) to[out=0, in=180] (mm.west);
    \draw[arrow] (mm) -- (z1);

    \node[font=\Large\bfseries] at (7, -3.4) {(a)};
\end{scope}

% Part (b)
\begin{scope}[yshift=-5cm, local bounding box=partb]
    \fill[gray!15, rounded corners] (-0.7,1.4) rectangle (14.8,-3);

    \node[font=\Large\itshape, align=center, anchor=north east] at (14.3, 1.3) {2 Independent parallelizable\\[-2pt]subgraphs};

    \node[input] (eps3) at (0, 0.7) {$\epsilon$};
    \node[input] (x3) at (0, -0.3) {$X$};
    \node[input] (y3) at (0, -2.3) {$W$};

    \node[rmsop] (sq3) at (2, 0) {square};
    \node[rmsop] (mean3) at (4, 0) {mean};
    \node[rmsop] (add3) at (6, 0) {add};
    \node[rmsop] (rsqrt3) at (8, 0) {rsqrt};

    \draw[dashbox, blue!70, thick] (1.0, 1.0) rectangle (9, -1);

    \node[matmulop] (mm3) at (6.5, -2.4) {matmul};
    \draw[dashbox, red!60, thick] (5.3, -1.8) rectangle (7.7, -3);

    \node[mulop] (mul3) at (12, -1.2) {multiply};
    \node[output] (z3) at (14, -1.2) {$Z$};

   \draw[->, line width=2pt] (7.8,2.5) -- (7.8,1);

    \draw[arrow] (x3.east) to[out=0, in=180] (sq3.west);
    \draw[arrow] (sq3) -- (mean3);
    \draw[arrow] (mean3) -- (add3);
    \draw[arrow] (eps3.east) to[out=10, in=160] (add3.north west);
    \draw[arrow] (add3) -- (rsqrt3);
    \draw[arrow] (rsqrt3.east) to[out=0, in=135] (mul3.north west);
    \draw[arrow] (x3.east) to[out=-60, in=120] (mm3.north);
    \draw[arrow] (y3.east) to[out=0, in=180] (mm3.west);
    \draw[arrow] (mm3.east) to[out=0, in=-135] (mul3.south west);
    \draw[arrow] (mul3) -- (z3);

    \node[font=\Large\bfseries] at (7, -3.4) {(b)};
\end{scope}
% Clip bounding box to remove extra whitespace
\pgfresetboundingbox
\useasboundingbox (-0.7,1.4) rectangle (15.8,-7.8);
\end{tikzpicture}}
\caption{(a) Computation graph for RMSNorm+MatMul (b) Semantically equivalent RMSNorm+MatMul graph with two independent subgraphs.}
\label{fig:rmsnorm-matmul-comp-graph}
\end{wrapfigure}
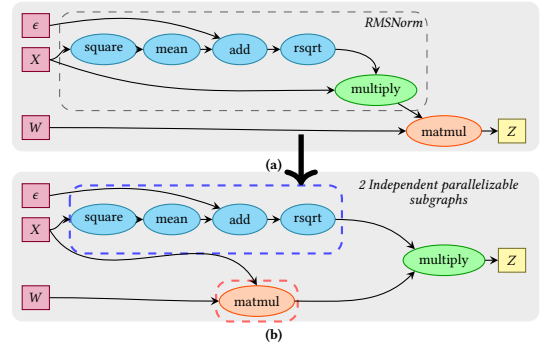

Figure~\ref{fig:rmsnorm-matmul-comp-graph}(a) shows the computation graph: the input $X$ is squared, reduced via \texttt{mean}, passed through \texttt{rsqrt}, multiplied with the original $X$, and finally given to \texttt{matmul} with weight matrix $W$.

\paragraph{Algebraic Transformations.}
On accelerators such as Trainium, executing this graph naively requires all operations to execute sequentially---the normalization must complete before the matrix multiplication can begin, underutilizing the available parallel compute engines.

\pname{} discovers that the element-wise \texttt{multiply} can be reordered with \texttt{matmul}: instead of computing $(X \odot \text{rms}(X)) \cdot W$, we can compute $(X \cdot W) \odot \text{rms}(X)$.
This is valid because $\text{rms}(X)$ reduces along the last dimension of $X$---the same dimension contracted by \texttt{matmul}---so the element-wise scaling commutes with the matrix multiplication.
\pname{} provably checks semantic equivalence by decomposing both operations into granular computations and using SMT over unbounded tensors (Section~\ref{sec:verification}), without requiring any hand-crafted rewrite rules.

The transformed graph in Figure~\ref{fig:rmsnorm-matmul-comp-graph}(b) exposes two independent subgraphs: \texttt{matmul} executes on the Tensor Engine while the RMS computation executes on the Scalar and Vector Engines in parallel (Section~\ref{sec:background}).
Rather than committing to a single transformation, \pname{} maintains both the original and transformed graphs simultaneously in a $\nu$Graph (``nu-graph'')---a persistent internal representation that tracks all semantically equivalent program variants through each stage of the synthesis pipeline (Section~\ref{sec:design}).

Previous works such as TASO~\cite{jia2019taso}, Tensat~\cite{tensat}, and Constable~\cite{vohra2025mind} use 91 hand-implemented rewrite rules\footnote{\url{https://github.com/uwplse/tensat/blob/master/src/rewrites.rs\#L12}} to perform graph rewrites. 
However, none of these works support this transformation. 
Although this particular rewrite could be added to the existing set of rules, implementing rewrite rules by hand for different combinations of tensor operators is error-prone, as demonstrated by recent efforts to formally verify such rules~\cite{arora2025tensorright}.
Mirage~\cite{wu2025mirage} does support this particular rewrite, but fails on a similar transformation in Softmax+MatMul (Appendix~\ref{app:softmax-matmul}). 
This is due to a fundamental limitation shared by all graph rewriting systems: they expect operations to appear in specified patterns and do not automatically reason about the semantics of operations to drive algebraic transformations.

\paragraph{Tiling.}
After discovering algebraic transformations, \pname{} tiles each tensor operation for the target hardware at multiple levels---from the granular units consumed by individual instructions up to the groups of data that reside in on-chip buffer simultaneously (Section~\ref{subsec:tiling}).
The transformed RMSNorm+MatMul graph now has two parallel branches, and \pname{} introduces \textit{independent} symbolic tiling parameters for each branch---enabling each to be optimized separately for its respective compute engine and on-chip buffer budget.
Different compute engines impose different constraints on tile dimensions, and the optimal grouping of tiles depends on the trade-off between data reuse and on-chip buffer capacity.
\pname{} generates multiple versions with different tiling configurations and selects the best empirically (Section~\ref{subsec:code-emitter}).

\paragraph{ISA Synthesis.}
\pname{} then lowers each tiled 2-D tensor operation to target ISA instructions via program synthesis.
For instance, on Trainium, the \texttt{matmul} operation must be lowered to the NKI instruction \texttt{nc\_matmul}, which computes $\mathbf{a}^\top \mathbf{b}$---the first operand is consumed in transposed form.
The synthesizer discovers that \texttt{matmul(x, y)} can be implemented as \texttt{nc\_matmul(nc\_transpose(x), y)}, and provably checks this using SMT-based equivalence checking (Section~\ref{sec:verification}).
For the element-wise operations in the RMS branch, the synthesizer explores multiple Trainium instruction choices---\texttt{tensor\_tensor}, \texttt{tensor\_scalar}, or \texttt{activation}---each with different performance characteristics and tile dimension constraints.
All valid instruction selections are maintained simultaneously in the $\nu$Graph (Section~\ref{sec:design}).

\paragraph{Fusion.}
After ISA synthesis, \pname{} identifies opportunities to fuse operations at two levels.
\textit{Node fusion} merges consecutive operations that share compatible tiling dimensions into a single loop nest, so that intermediate results remain in on-chip buffers rather than spilling to HBM.
For example, in the RMS branch, consecutive element-wise operations (such as \texttt{add\_epsilon} and \texttt{rsqrt}) can be fused to eliminate an HBM round-trip for the intermediate tensor.
\textit{Instruction fusion} replaces sequences of ISA instructions with fused variants: for example, on Trainium, a \texttt{dma\_copy} followed by \texttt{nc\_transpose} can be replaced with the NKI instruction \texttt{dma\_transpose}, performing the layout transformation on-the-fly during data transfer.
Additionally, ISA synthesis may introduce new operations (e.g., \texttt{nc\_transpose} on Trainium to satisfy \texttt{nc\_matmul}'s transposed-input requirement) that create new fusion opportunities not visible in the original graph.

\paragraph{Code Generation and Selection.}
Finally, \pname{} extracts concrete NKI programs from the $\nu$Graph by instantiating symbolic tiling parameters with specific values.
Each candidate program---encoding a specific combination of algebraic rewrite, instruction selection, tiling configuration, and fusion decisions---is compiled by the Neuron compiler and executed on Trainium.
\pname{} returns the best-performing kernel along with its configuration.
The full NKI kernel generated by \pname{} for this example is shown in Appendix~\ref{app:generated-nki}.

\section{Background}
\label{sec:background}

% AI accelerators differ in architectural details, but share common patterns: specialized compute engines (systolic arrays, vector and scalar units), a multi-level memory hierarchy with SRAM and HBM, and tile-based execution where programmers must manage data movement and tiling explicitly.
% \zsw{this is wrong, HBM is on-chip, and tile-based execution is not native to accelerators, and GPU does not have vector or scalar units.}
%\zswchanged{AI accelerators differ in architectural details, but they share common patterns, e.g., specialized units for computing matmuls, multi-level memory hierarchies, and requiring tiling of computations and tuning to achieve peak performance.}
AI accelerators differ in architectural details, but they share common patterns, e.g., specialized units for computing matmuls, multi-level memory hierarchies, and a reliance on tiling and tuning to achieve peak performance.
% AI accelerators share a common architectural pattern: specialized compute engines (systolic arrays, vector and scalar units), a multi-level memory hierarchy with on-chip SRAM and off-chip HBM, and tile-based execution where programmers must manage data movement and tiling explicitly.
%
Tile-based kernel languages exist for various accelerators---NKI for Amazon's Trainium~\cite{aws_neuron_nki_index}, Pallas for Google's TPU~\cite{pallas}, and cuTile~\cite{nvidiacutile} or Triton~\cite{tillet2019triton} for GPUs---that expose architectural features through Python-based interfaces.
This section introduces these concepts using Amazon's first-generation Trainium as a concrete example.

\begin{wrapfigure}{r}{0.4\textwidth}
\vspace{-20pt}
  \begin{center}
  \includegraphics[width=0.4\textwidth]{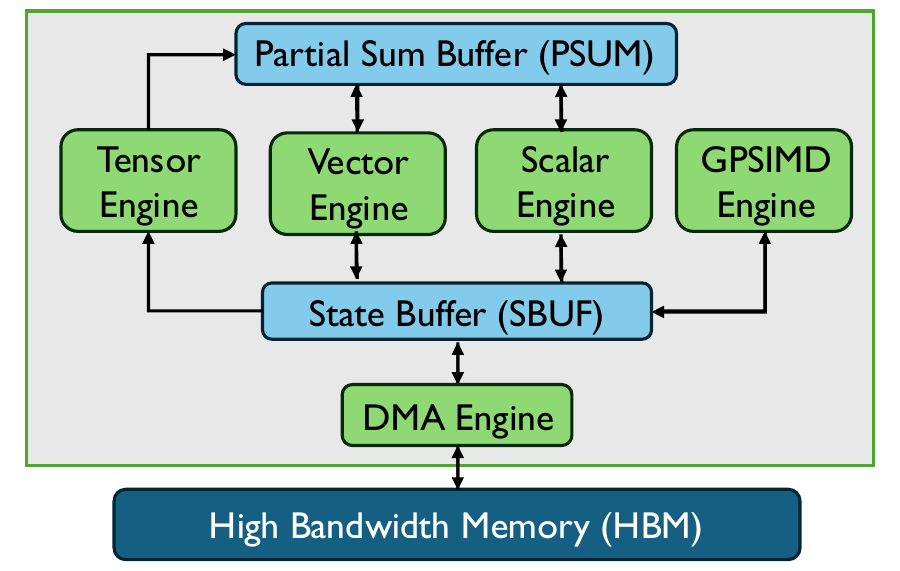}
  \end{center}
  \caption{Trainium's NeuronCore~\cite{aws_neuron_trainium}.}
\vspace{-10pt}
  \label{fig:neuron-arch}
\end{wrapfigure}

\subsection{Trainium Architecture}

Figure~\ref{fig:neuron-arch} illustrates the architecture of a NeuronCore, the fundamental compute unit in Trainium.
Each Trainium chip contains two NeuronCores.
A NeuronCore comprises four specialized compute engines: (1) a \textit{Tensor Engine} (systolic array) for matrix multiplications, (2) a \textit{Vector Engine} for vector operations, (3) a \textit{Scalar Engine} for scalar operations, and (4) a \textit{Gpsimd Engine}, a general-purpose SIMD unit.
All four engines can execute in parallel.

A NeuronCore has High Bandwidth Memory (HBM) and on-chip buffers: State Buffer (SBUF), the primary data buffer accessible by all compute engines, and Partial Sum Buffer (PSUM), an accumulator for the Tensor Engine. DMA engines move data between HBM and SBUF, and can operate in parallel with the compute engines.
%\zsw{HBM is usually considered on-chip since it is actually on the same chip as the core (though not the same silicon). Let us avoid this possibly confusing wording, and just use terms like DRAM/SRAM to be clear throughout.} ).
%

\subsection{Tile-Based Programming in NKI}

In tile-based programming, computation is structured around fixed-size sub-arrays called \textit{tiles} that are explicitly moved between HBM and on-chip SRAM buffers, with operations expressed at the tile granularity~\cite{tillet2019triton, wang2025tilelang}.
Neuron Kernel Interface (NKI)~\cite{aws_neuron_nki_index} is a Python-based tile-based language for implementing custom kernels on Trainium.
NKI kernels follow three stages: (1) loading tiles from HBM to SBUF via DMA, (2) performing on-chip computations on tiles, and (3) writing result tiles back from SBUF to HBM.

Figure~\ref{fig:nki-matmul} shows a tiled matrix multiplication kernel in NKI, adapted from the NKI tutorial~\cite{nki-matmul-tutorial}.
NKI kernels are Python functions decorated with \texttt{@nki.jit} (line 1).
The left-hand operand is provided pre-transposed (\texttt{lhsT}) because \texttt{nc\_matmul} consumes its stationary operand in transposed form.
The kernel iterates over output tiles (lines 11--12) and the reduction dimension (line 15), loading input tiles from HBM to SBUF via \texttt{dma\_copy} (lines 17--20), performing the matrix multiplication on the Tensor Engine via \texttt{nc\_matmul} which accumulates results into PSUM (line 23), and storing the result back to HBM via SBUF (lines 26--28).

Each compute engine enforces hardware constraints on tile dimensions: on Trainium, the partition dimension is at most 128 elements (\texttt{T\_K}, \texttt{T\_M}), while the free dimension of the moving operand can be up to 512 (\texttt{T\_N}) (lines 5--7).
Effective tiling maximizes on-chip data reuse and minimizes DMA traffic between HBM and SBUF.

\begin{wrapfigure}{r}{0.5\textwidth}
%\vspace{-25pt}
\scalebox{0.85}{\parbox{\textwidth}{\input{figures/nki_matmul_example.tex}}}
\caption{NKI kernel for tiled matrix multiplication on Trainium. The code iterates over tiles (lines 11--12), loads tiles from HBM to SBUF (lines 17--20), computes on-chip via \texttt{nc\_matmul} (line 23), and stores results back to HBM (lines 26--28). Tile dimensions (\texttt{TILE\_K}=128, \texttt{TILE\_N}=512) are constrained by the hardware.}
\label{fig:nki-matmul}
\vspace{-10pt}
\end{wrapfigure}

\subsection{Beyond Trainium}

%The architectural pattern described above---specialized compute and data movement engines, tile-based execution with hardware-constrained tile dimensions, and a multi-level memory hierarchy---is shared by other accelerators \cite{jouppi2023tpu, jouppi2017datacenter}.
%
%For example, Google's TPU exposes a similar programming model through Pallas~\cite{pallas}: programmers specify \texttt{BlockSpec} tile dimensions subject to hardware constraints (e.g., last two dimensions divisible by 8 and 128), computation executes on the MXU (matrix unit), vector, and scalar units, and data moves between HBM and on-chip VMEM.
%
%TileLang~\cite{wang2025tilelang} follows a similar tile-based pattern for GPUs.

Adapting \pname{} to target a new platform that implements a similar architecture would require providing new ISA semantics and hardware constraints, and modifying the code emitter to generate kernels for the new target; the core algorithms—algebraic transformations, tiling exploration, synthesis, and fusion—would remain unchanged.
%Adapting \pname{} to target a new platform such as Pallas would require providing new ISA semantics and hardware constraints for the TPU instruction set, and modifying the code emitter to generate Pallas kernels; the core algorithms---algebraic transformations, tiling exploration, synthesis, and fusion---would remain unchanged.
%
However, not all accelerator programming models are a good fit for \pname{}'s current design.
Many-core accelerators such as Meta's MTIA~\cite{coburn2025meta} and Microsoft's Maia~\cite{xu2024microsoftmaia} follow fundamentally different execution models without the specialized multi-engine architecture that \pname{} relies on.
GPU kernel languages such as Triton~\cite{tillet2019triton}, while tile-based, expose a thread-block execution model with shared memory and warp-level primitives that differ substantially from the DMA-driven, engine-parallel model of accelerators like Trainium.
Extending \pname{} to these platforms would require rethinking the ISA synthesis and fusion strategies to account for their distinct execution models.

\section{Design}
\label{sec:design}

This section describes the technical details of each stage in \pname{}'s synthesis pipeline.
We first introduce the $\nu$Graph (Section~\ref{subsec:nu-graphs}), the persistent internal representation that tracks all semantically equivalent program variants throughout the pipeline.
We then describe each pipeline stage: algebraic transformations via operator propagation (Section~\ref{subsec:tensor-graph-mutate}), symbolic tiling (Section~\ref{subsec:tiling}), ISA instruction synthesis (Section~\ref{subsec:synthesizer}), operator and instruction fusion (Section~\ref{subsec:isa-egraph-mutator}), and code emission with empirical selection (Section~\ref{subsec:code-emitter}).
We continue to use the RMSNorm+MatMul kernel introduced in Section~\ref{sec:motivation} as our running example.

\subsection{$\nu$Graph}
\label{subsec:nu-graphs}

A key challenge in superoptimization is maintaining multiple semantically equivalent program variants simultaneously---across algebraic reorderings, tiling configurations, instruction selections, and fusion choices---without committing to any single choice prematurely.
Equality saturation~\cite{willsey2021egg} addresses a similar problem using \textit{e-graphs}, which compactly represent a set of semantically equivalent program terms by grouping equivalent sub-expressions into \textit{e-classes}~\cite{vohra2025mind}.
\pname{} extends this idea to dataflow graphs of tensor operations with a structure we call a $\nu$Graph.

\begin{definition}[$\nu$Graph]
\label{def:nugraph}
A $\nu$Graph $\mathcal{G} = (\mathcal{V}, M)$ consists of:
\begin{itemize}
  \item $\mathcal{V} = \{G_0, G_1, \ldots, G_n\}$: a set of semantically equivalent dataflow graph variants, where each $G_i$ is a directed acyclic graph of tensor operations;
  \item $M$: metadata associated with each operation, which accumulates as the $\nu$Graph is lowered through the synthesis pipeline.
\end{itemize}
\end{definition}

Unlike standard e-graphs, which track equivalence classes of sub-expressions within a single program, a $\nu$Graph tracks equivalence classes of \textit{entire dataflow graphs}.
Each variant carries its own copy of metadata $M$: when a new variant is created (e.g., by an algebraic transformation or an ISA synthesis choice), it receives a fresh copy of the metadata with its own constraints.
The metadata $M$ evolves through the pipeline stages:
\begin{itemize}
  \item After \textit{algebraic transformations} (Section~\ref{subsec:tensor-graph-mutate}): multiple graph variants $\{G_0, \ldots, G_n\}$ representing different operator orderings.
  \item After \textit{tiling} (Section~\ref{subsec:tiling}): symbolic tiling parameters---strip dimensions $[n_0, n_1]$, block dimensions $[b_0, b_1]$, and tile dimensions $[t_0, t_1]$---for each operation.
  \item After \textit{ISA synthesis} (Section~\ref{subsec:synthesizer}): target ISA instructions, hardware constraints on tile dimensions, and compute engine assignments.
\end{itemize}

\begin{wrapfigure}{r}{0.5\textwidth}
\centering
\vspace{-20pt}
\raisebox{0pt}[\height][0pt]{\begin{tikzpicture}[
    scale=0.45,
    transform shape,
    input/.style={rectangle, draw=purple!70!black, fill=purple!30, minimum width=0.8cm, minimum height=0.7cm, font=\Large},
    output/.style={rectangle, draw=olive!70!black, fill=yellow!40, minimum width=0.8cm, minimum height=0.7cm, font=\Large},
    mulop/.style={ellipse, draw=green!60!black, fill=green!35, minimum width=1.6cm, minimum height=0.9cm, font=\Large},
    siluop/.style={ellipse, draw=teal!60!black, fill=teal!35, minimum width=1.3cm, minimum height=0.9cm, font=\Large},
    matmulop/.style={ellipse, draw=red!40!orange, fill=red!25!orange!30, minimum width=1.6cm, minimum height=0.9cm, font=\Large},
    recmatmulop/.style={rectangle, draw=red!40!orange, fill=red!25!orange!30, minimum width=1.6cm, minimum height=0.9cm, font=\Large},
    rectranspop/.style={rectangle, draw=yellow!40!orange, fill=yellow!25!orange!30, minimum width=1.6cm, minimum height=0.9cm, font=\Large},
    silughostop/.style={ellipse, draw=gray!60, dashed, fill=darkgray!20, minimum width=1.3cm, minimum height=0.9cm, font=\Large, text=black!70},
    mulghostop/.style={ellipse, draw=gray!60, dashed, fill=darkgray!20, minimum width=1.6cm, minimum height=0.9cm, font=\Large, text=black!70},
    arrow/.style={->, >=stealth},
    redarrow/.style={->, >=stealth, red!70, dashed}
]

% Part (a)
\begin{scope}[local bounding box=parta]
    \fill[gray!15, rounded corners] (-0.7,0) rectangle (14.8,-3);

    \node[input] (x) at (2, -0.5) {$X$};
    \node[input] (y) at (2, -2.5) {$Y$};
    \node[output] (z1) at (12, -1.5) {$Z$};

    \node[matmulop] (mm1) at (7, -1.5) {matmul};

    \draw[arrow] (x.east) to[out=0, in=150] (mm1.north west);
    \draw[arrow] (y.east) to[out=0, in=210] (mm1.south west);
    \draw[arrow] (mm1) -- (z1);

    \node[font=\Large\bfseries] at (7, -3.4) {(a)};
\end{scope}

% Part (b)
\begin{scope}[yshift=-4cm, local bounding box=partb]
    \fill[gray!15, rounded corners] (-0.7,0) rectangle (14.8,-3);

   \node[input] (x1) at (2, -0.5) {$X$};
    \node[input] (y1) at (2, -2.5) {$Y$};
    \node[output] (z2) at (12, -1.5) {$Z$};

    \node[recmatmulop, align=center] (mm2) at (7, -1.5) {[$n_0$, $n_1$, $b_0$, $b_1$] \\
    matmul $[t_0, t_1]$};

    \node[text=black!60!black,
    align=left, font=\fontsize{15}{12}\selectfont] at (9.7, 0.5) {\textit{Perform Tiling}};
    \draw[->, line width=2pt] (7.6,1.3) -- (7.6,-0.3);

   \draw[arrow] (x1.east) to[out=0, in=150] (mm2.north west);
    \draw[arrow] (y1.east) to[out=0, in=210] (mm2.south west);
    \draw[arrow] (mm2) -- (z2);

    \node[font=\Large\bfseries] at (7, -3.4) {(b)};
\end{scope}

% Part (c)
\begin{scope}[yshift=-12cm, local bounding box=partc]
    \fill[gray!15, rounded corners] (-0.7,4) rectangle (14.8,-5);

   \node[input] (x1) at (1, -2.5) {$X$};
    \node[input] (y1) at (1, -4) {$Y$};
    \node[output] (z2) at (14, -2.5) {$Z$};

    \node[input] (x2) at (1, 1.5) {$X$};
    \node[input] (y2) at (1, 0) {$Y$};
    \node[output] (z3) at (14, 1.5) {$Z$};

    \node[recmatmulop, align=center] (mm2) at (11, -2.5) {[$n_0$, $n_1$, $b_0$, $b_1$] \\
    nc\_matmul $[t_0, t_1]$};

    \node[rectranspop, align=center] (transp) at (5, -2.5) {[$n_0$, $n_1$, $b_0^{r}$, $b_1^{r}$] \\
    dma\_transpose $[t_0^{r}, t_1^{r}]$};

    \node[recmatmulop, align=center] (mm3) at (11, 1.5) {[$n_0$, $n_1$, $b_0$, $b_1$] \\
    nc\_matmul $[t_0, t_1]$};

    \node[rectranspop, align=center] (transp1) at (5, 1.5) {[$n_0$, $n_1$, $b_0^{r}$, $b_1^{r}$] \\
    nc\_transpose $[t_0^{r}, t_1^{r}]$};

    \node[font=\Large\itshape, align=center] at (5, 2.6) {Constraints: $t_0^{r}, t_1^{r}$ $\leq$ 128 \\
    Engine: Vector};

     \node[font=\Large\itshape, align=center] at (11, 2.6) {Constraints: $t_0$ $\leq$ 128, $t_1$ $\leq$ 512 \\
    Engine: Tensor};

    \node[font=\Large\itshape, align=center] at (5, -1.3) {Constraints: $t_0^{r}, t_1^{r}$ $\epsilon$ [128, 1024] \\
    Engine: DMA};

     \node[font=\Large\itshape, align=center] at (11, -1.3) {Constraints: $t_0$ $\leq$ 128, $t_1$ $\leq$ 512 \\
    Engine: Tensor};

    \node[text=black!60!black,
    align=left, font=\fontsize{15}{12}\selectfont] at (10.6, 4.5) {\textit{Synthesize Trainium ISA}};
    \draw[->, line width=2pt] (7.6,5.3) -- (7.6,3.6);

   \draw[arrow] (x1) -- (transp);
   \draw[arrow] (transp) -- (mm2);
    \draw[arrow] (y1.east) to[out=0, in=210] (mm2.south);
    \draw[arrow] (mm2) -- (z2);

    \draw[arrow] (x2) -- (transp1);
   \draw[arrow] (transp1) -- (mm3);
    \draw[arrow] (y2.east) to[out=0, in=210] (mm3.south);
    \draw[arrow] (mm3) -- (z3);

    \node[font=\Large\bfseries] at (7, -5.5) {(c)};
\end{scope}

% Reset and clip bounding box
\pgfresetboundingbox
\useasboundingbox (-0.7,1.4) rectangle (14.8,-17);

\end{tikzpicture}}
\caption{Example of lowering of $\nu$Graphs from a high-level computational graph in (a) to a hardware-agnostic tiled representation in (b), and to two variants of graphs with Trainium-specific ISA, and information on hardware-specific tiling constraints and engines in (c).}
\vspace{-10pt}
\label{fig:nugraph}
\end{wrapfigure}
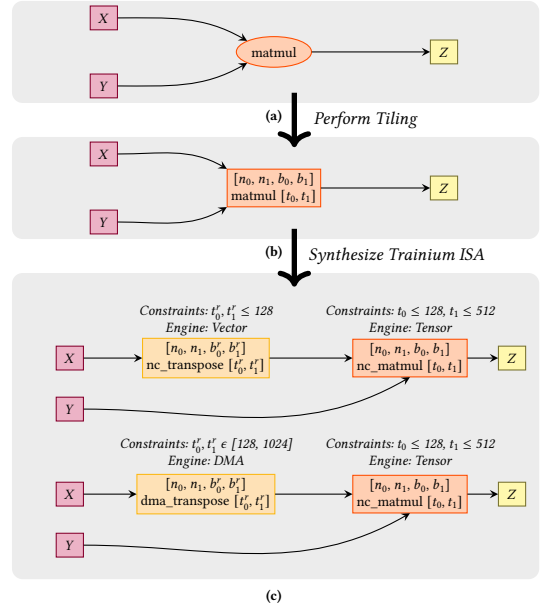

\paragraph{Exploration and extraction.}
Like equality saturation, the $\nu$Graph follows an \textit{exploration} phase and an \textit{extraction} phase.
During exploration, each pipeline stage may expand the set of variants: algebraic transformations generate alternative operator orderings, tiling introduces multiple block size configurations, and ISA synthesis discovers different instruction sequences for the same operation.
These choices compose multiplicatively---for example, 2 algebraic variants $\times$ 4 tiling configurations $\times$ 3 instruction choices yields 24 candidate programs.
During extraction (Section~\ref{subsec:code-emitter}), \pname{} instantiates the symbolic parameters, emits concrete programs from the $\nu$Graph, and selects the best-performing one through empirical evaluation on the target hardware.

\paragraph{Example.}
Figure~\ref{fig:nugraph} illustrates the $\nu$Graph progression on a matmul example.
In (a), the $\nu$Graph is a high-level dataflow graph with a single \texttt{matmul} operation.
In (b), after tiling, the operation is annotated with symbolic tiling parameters: $[n_0, n_1]$ for strip, $[b_0, b_1]$ for block dimensions, and $[t_0, t_1]$ for tile dimensions.
Note that this tiling strategy reflects the two-level memory hierarchy of AI accelerators such as Trainium (HBM and on-chip SRAM), where computations and data movement occur at tile granularity.
In (c), after ISA synthesis, the single \texttt{matmul} has been lowered to two variant instruction sequences---one using \texttt{nc\_transpose} on the Vector Engine (with tile constraints $t_0^r, t_1^r \leq 128$) and another using \texttt{dma\_transpose} on the DMA Engine (with tile constraints $t_0^r, t_1^r \in [128, 1024]$)---both followed by \texttt{nc\_matmul} on the Tensor Engine.
Both variants coexist in the $\nu$Graph with their respective hardware constraints; the best is selected empirically after code emission.

\paragraph{Benefits and limitations.}
By deferring all optimization choices to empirical selection, the $\nu$Graph avoids heuristic-based decisions that may be suboptimal for specific input shapes or hardware configurations.
It enables joint exploration across all optimization dimensions rather than making greedy sequential decisions that foreclose later opportunities.
However, because the number of variants grows multiplicatively across pipeline stages, \pname{} enumerates all valid combinations and relies on the target backend to reject infeasible ones (e.g., programs that exceed on-chip buffer capacity).
In practice, this approach is tractable for kernel-sized programs: ML kernels typically contain fewer than 10 operators per layer, and the resulting search space (up to ${\sim}$30K candidates for complex kernels such as GQA and Gated MLP) can be evaluated within two hours as a one-time cost (Section~\ref{subsec:comp-times}).

\begin{wrapfigure}{r}{0.5\textwidth}
\centering
\vspace{-20pt}

\raisebox{0pt}[\height][0pt]{\begin{tikzpicture}[
    scale=0.45,
    transform shape,
    input/.style={rectangle, draw=purple!70!black, fill=purple!30, minimum width=0.8cm, minimum height=0.7cm, font=\Large},
    output/.style={rectangle, draw=olive!70!black, fill=yellow!40, minimum width=0.8cm, minimum height=0.7cm, font=\Large},
    mulop/.style={ellipse, draw=green!60!black, fill=green!35, minimum width=1.6cm, minimum height=0.9cm, font=\Large},
    siluop/.style={ellipse, draw=teal!60!black, fill=teal!35, minimum width=1.3cm, minimum height=0.9cm, font=\Large},
    matmulop/.style={ellipse, draw=red!40!orange, fill=red!25!orange!30, minimum width=1.6cm, minimum height=0.9cm, font=\Large},
    silughostop/.style={ellipse, draw=gray!60, dashed, fill=darkgray!20, minimum width=1.3cm, minimum height=0.9cm, font=\Large, text=black!70},
    mulghostop/.style={ellipse, draw=gray!60, dashed, fill=darkgray!20, minimum width=1.6cm, minimum height=0.9cm, font=\Large, text=black!70},
    arrow/.style={->, >=stealth},
    redarrow/.style={->, >=stealth, red!70, dashed}
]

% Part (a)
\begin{scope}[local bounding box=parta]
    \fill[gray!15, rounded corners] (-0.7,1.4) rectangle (14.8,-3);

    \node[input] (x1) at (0, 0.7) {$X$};
    \node[input] (w1) at (0, -0.3) {$W_1$};
    \node[input] (w2) at (0, -1.3) {$W_2$};
    \node[input] (w3) at (0, -2.3) {$W_3$};

    \node[matmulop] (mm1) at (4, 0) {matmul};
    \node[matmulop] (mm2) at (2.5, -1.3) {matmul};
    \node[siluop] (silu) at (6.3, 0) {silu};
    \node[mulop] (mul) at (9, 0) {multiply};
    \node[matmulop] (mm3) at (12, -1) {matmul};
    \node[output] (z1) at (14, -1) {$Z$};

    \draw[arrow] (x1.east) to[out=0, in=90] (mm1.north);
    \draw[arrow] (w1.east) to[out=0, in=180] (mm1.west);
    \draw[arrow] (x1.east) to[out=0, in=90] (mm2.north);
    \draw[arrow] (w2.east) to[out=0, in=180] (mm2.west);
    \draw[arrow] (mm1) -- (silu);
    \draw[arrow] (silu) -- (mul);
    \draw[arrow] (mm3) -- (z1);
    \draw[arrow] (mm2.east) to[out=0, in=220] (mul.south);
    \draw[arrow] (mul.east) to[out=0, in=150] (mm3.north);
    \draw[arrow] (w3.east) to[out=0, in=200] (mm3.west);

    \node[font=\Large\bfseries] at (7, -3.4) {(a)};
\end{scope}

% Part (b)
\begin{scope}[yshift=-5.5cm, local bounding box=partb]
    \fill[gray!15, rounded corners] (-0.7,1.4) rectangle (14.8,-3);

    \node[input] (x1) at (0, 0.7) {$X$};
    \node[input] (w1) at (0, -0.3) {$W_1$};
    \node[input] (w2) at (0, -1.3) {$W_2$};
    \node[input] (w3) at (0, -2.3) {$W_3$};

    \node[matmulop] (mm1) at (4, 0) {matmul};
    \node[matmulop] (mm2) at (2.5, -1.3) {matmul};
    \node[siluop] (silu) at (6.3, 0) {silu};
    \node[mulop] (mul) at (9, 0) {multiply};
    \node[silughostop] (silunew) at (12, 0.8) {silu};
    \node[matmulop] (mm3) at (12, -1) {matmul};
    \node[output] (z1) at (14, -1) {$Z$};
    \node[text=red!60!black, font=\fontsize{50}{38}\selectfont] at (13.5, 0.8) {$\times$};

    \draw[arrow] (x1.east) to[out=0, in=90] (mm1.north);
    \draw[arrow] (w1.east) to[out=0, in=180] (mm1.west);
    \draw[arrow] (x1.east) to[out=0, in=90] (mm2.north);
    \draw[arrow] (w2.east) to[out=0, in=180] (mm2.west);
    \draw[arrow] (mm1) -- (silu);
    \draw[arrow] (silu) -- (mul);
    \draw[arrow] (mm3) -- (z1);
    \draw[arrow] (mm2.east) to[out=0, in=220] (mul.south);
    \draw[arrow] (mul.east) to[out=0, in=150] (mm3.north);
    \draw[arrow] (w3.east) to[out=0, in=200] (mm3.west);
    \draw[redarrow,  thick] (mul.north) to[out=90, in=180] (silunew.west);
    \draw[redarrow,  thick] (silunew.south) to[out=270, in=90] (mm3.north);
    \draw[redarrow,  thick] (mm1.south) to[out=320, in=190] (mul.south);

    \node[font=\Large\bfseries] at (7, -3.4) {(b)};
\end{scope}

% Part (c)
\begin{scope}[yshift=-11cm, local bounding box=partc]
    \fill[gray!15, rounded corners] (-0.7,1.4) rectangle (14.8,-3);

    \node[font=\Large\itshape, align=center, anchor=north east] at (14.3, 1.3) {};

    \node[input] (x1) at (0, 0.7) {$X$};
    \node[input] (w1) at (0, -0.3) {$W_1$};
    \node[input] (w2) at (0, -1.3) {$W_2$};
    \node[input] (w3) at (0, -2.3) {$W_3$};

    \node[matmulop] (mm1) at (4, 0) {matmul};
    \node[matmulop] (mm2) at (2.5, -1.3) {matmul};
    \node[siluop] (silu) at (6.3, 0) {silu};
    \node[mulop] (mul) at (9, 0) {multiply};
    \node[mulghostop] (mulnew) at (12, 0.8) {multiply};
    \node[matmulop] (mm3) at (12, -1) {matmul};
    \node[output] (z1) at (14, -1) {$Z$};
    \node[text=red!60!black, font=\fontsize{50}{38}\selectfont] at (14.4, 0.8) {$\times$};

    \draw[arrow] (x1.east) to[out=0, in=90] (mm1.north);
    \draw[arrow] (w1.east) to[out=0, in=180] (mm1.west);
    \draw[arrow] (x1.east) to[out=0, in=90] (mm2.north);
    \draw[arrow] (w2.east) to[out=0, in=180] (mm2.west);
    \draw[arrow] (mm1) -- (silu);
    \draw[arrow] (silu) -- (mul);
    \draw[arrow] (mm3) -- (z1);
    \draw[arrow] (mm2.east) to[out=0, in=220] (mul.south);
    \draw[arrow] (mul.east) to[out=0, in=150] (mm3.north);
    \draw[arrow] (w3.east) to[out=0, in=200] (mm3.west);
    \draw[redarrow,  thick] (mm3.north) to[out=90, in=270] (mulnew.south);
    \draw[redarrow,  thick] (mulnew.east) to[out=0, in=90] (z1.north);
    \draw[redarrow,  thick] (silu.south) to[out=320, in=270] (mm3.south);

    \node[font=\Large\bfseries] at (7, -3.4) {(c)};
\end{scope}

% Reset and clip bounding box
\pgfresetboundingbox
\useasboundingbox (-0.7,1.4) rectangle (14.8,-14);

\end{tikzpicture}}

\caption{Gated MLP: (a) Original computation graph; (b) and (c) show invalid attempts to propagate \texttt{silu} and \texttt{multiply} operations (dotted red edges). Since no valid transformation exists, no new $\nu$Graph is generated.}
\vspace{-20pt}
\label{fig:mlp-example}
\end{wrapfigure}
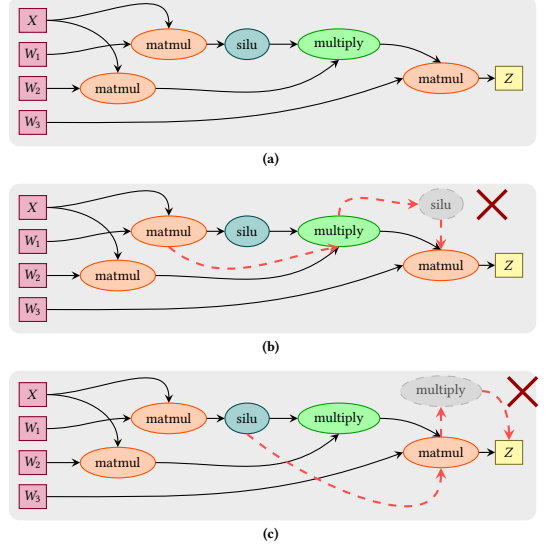

\subsection{Tensor Computation Graph Mutator}
\label{subsec:tensor-graph-mutate}

The Tensor Computation Graph Mutator takes a tensor computation graph as input and performs \textit{provably correct}, graph-level algebraic transformations to enable better parallelization, ISA fusion, and loop fusion opportunities.
As illustrated in Section~\ref{sec:motivation}, the key insight is that certain operations can be \textit{reordered} while preserving semantic equivalence---for example, propagating the \texttt{multiply} operation past \texttt{matmul} in the RMSNorm+MatMul kernel enables parallel execution of independent subgraphs.
Since multiple valid transformations may exist, each exposing different optimization opportunities, \pname{} maintains all discovered variants of semantically equivalent $\nu$Graphs $\{G_0, G_1, \ldots, G_n\}$.

\begin{wrapfigure}{r}{0.5\textwidth}
\vspace{-30pt}
\centering
\raisebox{0pt}[\height][0pt]{\begin{tikzpicture}[
    scale=0.45,
    transform shape,
    input/.style={rectangle, draw=purple!70!black, fill=purple!30, minimum width=0.8cm, minimum height=0.7cm, font=\Large},
    output/.style={rectangle, draw=olive!70!black, fill=yellow!40, minimum width=0.8cm, minimum height=0.7cm, font=\Large},
    mulop/.style={ellipse, draw=green!60!black, fill=green!35, minimum width=1.6cm, minimum height=0.9cm, font=\Large},
    siluop/.style={ellipse, draw=teal!60!black, fill=teal!35, minimum width=1.3cm, minimum height=0.9cm, font=\Large},
    matmulop/.style={ellipse, draw=red!40!orange, fill=red!25!orange!30, minimum width=1.6cm, minimum height=0.9cm, font=\Large},
    recmulop/.style={rectangle, draw=green!60!black, fill=green!35, minimum width=1.6cm, minimum height=0.9cm, font=\Large},
    recsiluop/.style={rectangle, draw=teal!60!black, fill=teal!35, minimum width=1.3cm, minimum height=0.9cm, font=\Large},
    recmatmulop/.style={rectangle, draw=red!40!orange, fill=red!25!orange!30, minimum width=1.6cm, minimum height=0.9cm, font=\Large},
    arrow/.style={->, >=stealth},
    redarrow/.style={->, >=stealth, red!70, dashed}
]

% Part (a)
\begin{scope}[local bounding box=parta]
    \fill[gray!15, rounded corners] (-0.7,1.4) rectangle (14.8,-3);

    \node[input] (x1) at (0, 0.7) {$X$};
    \node[input] (w1) at (0, -0.3) {$W_1$};
    \node[input] (w2) at (0, -1.3) {$W_2$};
    \node[input] (w3) at (0, -2.3) {$W_3$};

    \node[matmulop] (mm1) at (4, 0) {matmul};
    \node[matmulop] (mm2) at (2.5, -1.3) {matmul};
    \node[siluop] (silu) at (6.3, 0) {silu};
    \node[mulop] (mul) at (9, 0) {multiply};
    \node[matmulop] (mm3) at (12, -1) {matmul};
    \node[output] (z1) at (14, -1) {$Z$};

    \draw[arrow] (x1.east) to[out=0, in=90] (mm1.north);
    \draw[arrow] (w1.east) to[out=0, in=180] (mm1.west);
    \draw[arrow] (x1.east) to[out=0, in=90] (mm2.north);
    \draw[arrow] (w2.east) to[out=0, in=180] (mm2.west);
    \draw[arrow] (mm1) -- (silu);
    \draw[arrow] (silu) -- (mul);
    \draw[arrow] (mm3) -- (z1);
    \draw[arrow] (mm2.east) to[out=0, in=220] (mul.south);
    \draw[arrow] (mul.east) to[out=0, in=150] (mm3.north);
    \draw[arrow] (w3.east) to[out=0, in=200] (mm3.west);

    \node[font=\Large\bfseries] at (7, -3.4) {(a)};
\end{scope}

% Part (b)
\begin{scope}[yshift=-5.5cm, local bounding box=partb]
    \fill[gray!15, rounded corners] (-0.7,1.4) rectangle (14.8,-3);

    \node[input] (x1) at (0, 0.7) {$X$};
    \node[input] (w1) at (0, -0.3) {$W_1$};
    \node[input] (w2) at (0, -1.3) {$W_2$};
    \node[input] (w3) at (0, -2.3) {$W_3$};

    \node[recmatmulop, align=center] (mm1) at (4.5, 0) {[$n_0$, $n_1$, $b_0^{m_0}$, $b_1^{m_0}$] \\
    matmul $[t_0^{m_0}, t_1^{m_0}]$};
    \node[recmatmulop, align=center] (mm2) at (2.5, -1.3)
  {[$n_0$, $n_1$, $b_0^{m_1}$, $b_1^{m_1}$] \\
   matmul $[t_0^{m_1}, t_1^{m_1}]$};
    \node[recsiluop, align=center] (silu) at (8, 0) {[$n_0$, $n_1$, $b_0^{s}$, $b_1^{s}$] \\
   silu $[t_0^{s}, t_1^{s}]$};
    \node[recmulop, align=center] (mul) at (11.5, 0) {[$n_0$, $n_1$, $b_0^{m}$, $b_1^{m}$] \\
   multiply $[t_0^{m}, t_1^{m}]$};
    \node[recmatmulop, align=center] (mm3) at (11.5, -2) {[$n_0$, $n_1$, $b_0^{m_2}$, $b_1^{m_2}$] \\
   matmul $[t_0^{m_2}, t_1^{m_2}]$};
    \node[output] (z1) at (14, -2) {$Z$};

    \node[text=black!60!black,
    align=left, font=\fontsize{15}{12}\selectfont] at (11.5, 2) {Tiling strategy:
    \\
    $\forall$ op $\epsilon$ G:
    \\
    op $\rightarrow$ [$n_0$, $n_1$, $b_0$, $b_1$] op $[t_0, t_1]$};
    \draw[->, line width=2pt] (7.6,3) -- (7.6,1);

    \draw[arrow] (x1.east) to[out=0, in=90] (mm1.north);
    \draw[arrow] (w1.east) to[out=0, in=180] (mm1.west);
    \draw[arrow] (x1.east) to[out=0, in=90] (mm2.north);
    \draw[arrow] (w2.east) to[out=0, in=180] (mm2.west);
    \draw[arrow] (mm1) -- (silu);
    \draw[arrow] (silu) -- (mul);
    \draw[arrow] (mm3) -- (z1);
    \draw[arrow] (mm2.east) to[out=0, in=200] (mul.south);
    \draw[arrow] (mul.south) to[out=0, in=90] (mm3.north);
    \draw[arrow] (w3.east) to[out=0, in=200] (mm3.west);

    \node[font=\Large\bfseries] at (7, -3.4) {(b)};
\end{scope}

% Reset and clip bounding box
\pgfresetboundingbox
\useasboundingbox (-0.7,3.2) rectangle (14.8,-9);

\end{tikzpicture}}
\caption{Tiling of $\nu$Graph for Gated MLP with strip, block,
and tile dimensions for each operation.}
\vspace{-10pt}
\label{fig:tiled-mlp}
\end{wrapfigure}
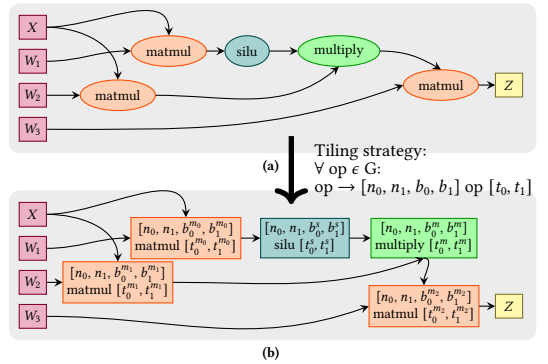

\begin{definition}[Semantic Equivalence]
\label{def:semantic-equiv}
Two computation graphs $G_1$ and $G_2$ are \emph{semantically equivalent}, denoted $G_1 \equiv G_2$, if for all input tensors satisfying the operator preconditions (Section~\ref{subsubsec:tensor-ops-sema}), both graphs produce identical output tensors.
\end{definition}

\pname{} preserves all variants where $G_0 \equiv G_1 \equiv \cdots \equiv G_n$ and selects the best-performing one empirically when concrete input shapes are known (Section~\ref{subsec:code-emitter}). 
Although the number of variants is worst-case exponential, in practice most operator pairs are not swappable and ML kernels typically contain < 10 operators.
%By preserving all variants where $G_0 \equiv G_1 \equiv \cdots \equiv G_n$, \pname{} defers the selection of the best-performing variant until concrete input shapes are known and empirical evaluation can be performed (Section~\ref{subsec:code-emitter}).
%
%While the number of variants is worst-case exponential in the number of operators, in practice most operator pairs are not swappable, and ML kernels typically contain < 10 operators (Section~\ref{subsec:nu-graphs}).

\paragraph{Operator Propagation.}
To discover algebraic transformations, \pname{} uses a worklist-based algorithm that attempts to \textit{propagate} each operation downward through the computation graph by swapping it with its successors.
For each operator, the algorithm repeatedly checks whether swapping it with an adjacent successor preserves semantic equivalence (Section~\ref{sec:verification}); if so, the new graph variant is added to the $\nu$Graph and propagation continues from the new position.
The process saturates when no further swaps are possible for any operator, ensuring that all reachable reorderings are explored.
Note that by pushing operators down, other operators are inevitably moved up; therefore, upward propagation is unnecessary.
Algorithm~\ref{alg:op-propagate} formalizes this process.

\begin{algorithm}[t]
 \scriptsize
 \caption{Generation of $\nu$Graph Variants}
 \label{alg:op-propagate}
 \begin{algorithmic}[1]
 \Require Computation graph $G_0$
 \State $\mathcal{M} \gets \{G_0\}$ \Comment{set of semantically-equivalent graph variants}
 \ForAll{$op_1$ in $G_0$} \Comment{try propagating each operator}
   \State $\mathcal{M}_{\text{next}} \gets \varnothing$
   \ForAll{$G \in \mathcal{M}$}
     \State $\mathcal{W} \gets \{(G, \text{pos of } op_1 \text{ in } G)\}$ \Comment{worklist of (graph, position)}
     \While{$\mathcal{W}\neq\varnothing$}
       \State $(G_{\text{cur}}, pos) \gets \mathcal{W}.pop()$; \, $advanced \gets \textbf{false}$
       \ForAll{immediate successor $op_2$ of $pos$ in $G_{\text{cur}}$}
         \State $\mathcal{I} \gets \textsf{GetInputs}(G_{\text{cur}}, pos)$ ; \, $accepted \gets \textbf{false}$
         \For{$k=1$ to $|\mathcal{I}|$} \Comment{progressively increase clone degree}
           \ForAll{$S \subseteq \mathcal{I}$ with $|S|=k$}
             \State $(G_{\text{new}}, p_{\text{new}}) \gets \textsf{SwapWithSuccessor}(G_{\text{cur}}, pos, op_2, S)$
             \If{$G_{\text{new}} \neq \textbf{null}$ \textbf{and} $\textsf{CheckEquivalence}(op_1, op_2, G_{\text{cur}}, G_{\text{new}})$}
               \State $\mathcal{M}_{\text{next}} \gets \mathcal{M}_{\text{next}} \cup \{G_{\text{new}}\}$ \Comment{save new variant}
               \State $\mathcal{W} \gets \mathcal{W} \cup \{(G_{\text{new}}, p_{\text{new}})\}$ \Comment{continue propagating}
               \State $advanced \gets \textbf{true}$ ; \, $accepted \gets \textbf{true}$
             \EndIf
           \EndFor
           \If{$accepted$}
             \State \textbf{break} \Comment{stop once equivalent swap found}
           \EndIf
         \EndFor
       \EndFor
       \If{$\neg advanced$}
         \State $\mathcal{M}_{\text{next}} \gets \mathcal{M}_{\text{next}} \cup \{G_{\text{cur}}\}$ \Comment{keep if no swap possible}
       \EndIf
     \EndWhile
   \EndFor
   \State $\mathcal{M} \gets \mathcal{M} \cup \mathcal{M}_{\text{next}}$
 \EndFor
 \State \Return $\mathcal{M}$
 \end{algorithmic}
 %\vspace{-.05in}
\end{algorithm}

Algorithm~\ref{alg:op-propagate} generates the $\nu$Graph variants as follows.
Starting with the original computation graph $G_0$ (line 1), the algorithm iterates over each operator $op_1$ in topological order and attempts to propagate it downward (line 2).
For each existing graph variant, it initializes a worklist with the current position of $op_1$ (line 5).
The algorithm repeatedly pops a (graph, position) pair from the worklist (line 7) and examines each immediate successor $op_2$ (line 8).
For each successor, it retrieves the input operands $\mathcal{I}$ of $op_1$ (line 9) and attempts swaps with progressively increasing \emph{clone degree} $k$ (lines 10--11): $k$ determines how many of $op_1$'s inputs pass through $op_2$ in the swap, preferring the smallest $k$ that yields a valid transformation (lines 17--18).
For each subset $S \subseteq \mathcal{I}$ of size $k$, $\textsf{SwapWithSuccessor}$ constructs the swapped graph, and $\textsf{CheckEquivalence}$ provably checks whether the swap preserves semantic equivalence (Section~\ref{sec:verification}).
The algorithm prefers the smallest clone degree that yields a valid swap (lines 17--18).
If the check passes, the new variant is saved and pushed onto the worklist to continue propagating (lines 13--15); if no swap is possible, the current graph is retained (lines 19--20).
After processing all operators, the algorithm returns the complete set of semantically equivalent graph variants.
%For each successor, it attempts swaps with progressively increasing clone degrees (lines 10--15): $\textsf{SwapWithSuccessor}$ constructs the swapped graph, and $\textsf{CheckEquivalence}$ provably checks whether the swap preserves semantic equivalence (Section~\ref{sec:verification}).
%
%If the check passes, the new variant is added to $\mathcal{M}_{\text{next}}$ and pushed onto the worklist to continue propagating (lines 13--14).
%
%If no swap is possible, the current graph is retained as-is (line 18).
%
%After processing all operators, the algorithm returns the complete set of semantically equivalent graph variants.

For example, in Figure~\ref{fig:rmsnorm-matmul-comp-graph}, the algorithm discovers that \texttt{multiply} can be propagated past \texttt{matmul}, producing the transformed graph in (b).
In contrast, Figure~\ref{fig:mlp-example} shows the Gated MLP kernel where propagation attempts fail the equivalence check, so no new variants are generated.

\subsubsection{Semantics of Tensor Operations}
\label{subsubsec:tensor-ops-sema}

Programmers provide operational semantics for the high-level tensor operations they use in their programs, as shown in Figure~\ref{fig:op-sema} for the \texttt{matmul} operation.
The semantics specification includes three key components:

\begin{itemize}
    \item \textit{\textbf{Rank and shape preconditions:}} Requirements on tensor ranks and dimension compatibility for a given tensor operation. In the example, \texttt{a.min\_rank() == 2} requires that the input tensors have at least two dimensions, and \texttt{a.shape[-1] == b.shape[-2]} ensures the contraction dimensions of the input tensors match.

    \item \textit{\textbf{Access maps:}} How input tensors are indexed, defined via symbolic axes (akin to affine maps in MLIR~\cite{lattner2021mlir}). The expression \texttt{a[..., i0, i1]} uses NumPy-like indexing style to denote accessing an element at given indices.

    \item \textit{\textbf{Granular semantics:}} The element-level operations expressed using granular operators. The expression \texttt{mul(a[..., i0, i1], b[..., i1, i2]).sum(axis=i1)} specifies multiply-accumulate over the contraction axis \texttt{i1}. These expressions are converted to SMT formulas used for equivalence checking (Section~\ref{sec:verification}).
\end{itemize}

This allows programmers to define custom high-level, hardware-agnostic tensor operations.

\subsection{Tensor Graph Tiling Pass}
\label{subsec:tiling}
The Tiling Pass prepares the Tensor Computation $\nu$Graph for ISA synthesis by introducing symbolic tiling parameters.
This transformation tiles each operation with symbolic strip, block, and tile dimension sizes, and converts N-D tensor operations into 2-D operations enclosed in loops.
Loops are not explicitly represented; instead, we use a compact representation showing symbolic dimensions.

\paragraph{Tiling Notation.}
Figure~\ref{fig:tiled-mlp} illustrates the tiling notation using the Gated MLP example, where $x$ is the input tensor and $w_1$, $w_2$, $w_3$ are weight matrices.
Each operation is annotated with $[n_0, n_1, b_0, b_1]\ op\ [t_0, t_1]$, where:
\begin{itemize}
  \item $[t_0, t_1]$: \textbf{Tile dimensions}---the size of each 2-D tile processed by a single instruction for a compute engine. These are determined by hardware constraints of the target instructions (described in Section~\ref{subsec:synthesizer}).

  \item $[b_0, b_1]$: \textbf{Block dimensions}---the number of tiles grouped into a block along each dimension.
  Blocking enables data reuse by keeping data in on-chip SRAM (SBUF/PSUM) across multiple tile operations; larger blocks increase reuse but are constrained by the capacity of on-chip SRAM. Block dimension size selection also impacts DMA utilization, as the DMA engine can move multiple tiles between HBM and SBUF; therefore, small block sizes can lead to DMA bandwidth underutilization.
  %\zswchanged{Blocking serves two main purposes for NKI: moving data in larger chunks between DRAM and SRAM can reduce DMA overhead; it can also help increase arithmetic intensity of the algorithm by increasing reuse of data to avoid getting bounded by memory bandwidth. However, block size is also constrained by the available SRAM on-chip.}

  \item $[n_0, n_1]$: \textbf{Strip dimensions}---the number of blocks along each dimension in HBM, computed as $\bigl[\ceil{\tfrac{M}{b_0 t_0}},\ \ceil{\tfrac{N}{b_1 t_1}}\bigr]$ for a tensor of shape $[M, N]$.

\end{itemize}

\begin{wrapfigure}{r}{0.5\textwidth}
\vspace{-10pt}
\begin{lstlisting}[style=pythonstyle]
@semantics()
def matmul (a : SymTensor, b : SymTensor):
 # Rank and shape preconditions
 precondition(a.min_rank()==2)
 precondition(b.min_rank()==2)
 precondition(a.shape[-1]==b.shape[-2])
 M,K,N = a.shape[-2],a.shape[-1],b.shape[-1]

 sema=OpSema(name="matmul")
 sema.operands=[a, b]
 # Access maps
 i0 = SymAxis("i0",range=[0,M])
 i1 = SymAxis("i1",range=[0,K],is_contracting=True)
 i2 = SymAxis("i2",range=[0,N])
 # Computation: a @ b
 sema.semantics = mul(a[..., i0, i1], b[..., i1, i2]).sum(axis=i1)
 return sema
\end{lstlisting}
%\vspace{-5pt}
\caption{Semantics of \texttt{matmul} tensor operation in \pname{}.}
\label{fig:op-sema}
\vspace{-10pt}
\end{wrapfigure}

For example, in Figure~\ref{fig:tiled-mlp}, the first \texttt{matmul} operation ($X \cdot W_1$) has block dimensions $[b_0^{m0}, b_1^{m0}]$ and tile dimensions $[t_0^{m0}, t_1^{m0}]$, where the superscript $m0$ identifies these as the first matmul's parameters.
Similarly, the \texttt{silu} operation has its own parameters $[b_0^{s}, b_1^{s}]$ and $[t_0^{s}, t_1^{s}]$, and subsequent matmuls use superscripts $m1$, $m2$.
Since the optimal block size depends on the trade-off between data reuse and on-chip buffer capacity, \pname{} uses symbolic parameters and generates multiple versions with block sizes ranging from 1 to 32 tiles (bounded by SBUF capacity on Trainium), selecting the best-performing configuration empirically (Section~\ref{subsec:code-emitter}).
%
%On Trainium, the upper bound of 32 tiles per block is determined by the physical capacity of the on-chip SBUF: blocks larger than 32 tiles exceed the available on-chip buffer for all benchmarks in our evaluation, so exploring larger block sizes would only produce infeasible candidates.
%
Strip dimensions along which no reduction occurs are shared across all tensor operators in a computation graph ($[n_0, n_1]$) to maximize data reuse of on-chip blocks across operators and facilitate fusion opportunities later in the synthesis pipeline.

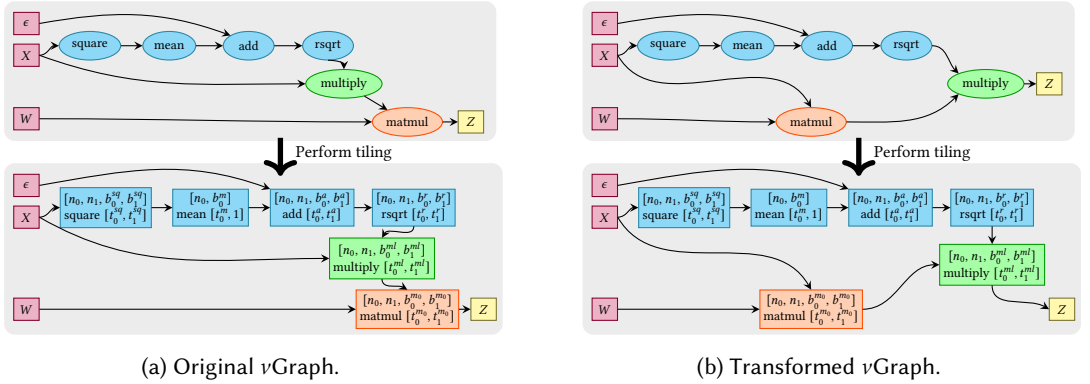
\begin{figure*}
\centering
\begin{minipage}[t]{0.45\textwidth}
  \centering
  \begin{tikzpicture}[
    scale=0.42,
    transform shape,
    input/.style={rectangle, draw=purple!70!black, fill=purple!30, minimum width=0.8cm, minimum height=0.7cm, font=\Large},
    output/.style={rectangle, draw=olive!70!black, fill=yellow!40, minimum width=0.8cm, minimum height=0.7cm, font=\Large},
    rmsop/.style={ellipse, draw=cyan!60!black, fill=cyan!40, minimum width=1.6cm, minimum height=0.9cm, font=\Large},
    mulop/.style={ellipse, draw=green!60!black, fill=green!35, minimum width=1.6cm, minimum height=0.9cm, font=\Large},
    matmulop/.style={ellipse, draw=red!40!orange, fill=red!25!orange!30, minimum width=1.6cm, minimum height=0.9cm, font=\Large},
    recrmsop/.style={rectangle, draw=cyan!60!black, fill=cyan!40, minimum width=1.6cm, minimum height=0.9cm, font=\Large},
    recmulop/.style={rectangle, draw=green!60!black, fill=green!35, minimum width=1.6cm, minimum height=0.9cm, font=\Large},
    recmatmulop/.style={rectangle, draw=red!40!orange, fill=red!25!orange!30, minimum width=1.6cm, minimum height=0.9cm, font=\Large},
    recrmsop/.style={rectangle, draw=cyan!60!black, fill=cyan!40, minimum width=1.6cm, minimum height=0.9cm, font=\Large},
    arrow/.style={->, >=stealth},
]

% Original graph
\begin{scope}[local bounding box=parta]
    \fill[gray!15, rounded corners] (-0.7,1.4) rectangle (14.8,-3);

    \node[input] (eps) at (0, 0.7) {$\epsilon$};
    \node[input] (x1) at (0, -0.3) {$X$};
    \node[input] (y1) at (0, -2.3) {$W$};

    \node[rmsop] (sq) at (2, 0) {square};
    \node[rmsop] (mean) at (4.5, 0) {mean};
    \node[rmsop] (add) at (7, 0) {add};
    \node[rmsop] (rsqrt) at (9.5, 0) {rsqrt};

    \node[mulop] (mul) at (10, -1.2) {multiply};
    \node[matmulop] (mm) at (12, -2.4) {matmul};
    \node[output] (z1) at (14, -2.4) {$Z$};

    \draw[arrow] (x1.east) to[out=0, in=180] (sq.west);
    \draw[arrow] (sq) -- (mean);
    \draw[arrow] (mean) -- (add);
    \draw[arrow] (eps.east) to[out=10, in=160] (add.north west);
    \draw[arrow] (add) -- (rsqrt);
    \draw[arrow] (rsqrt.south) to[out=0, in=90] (mul.north);
    \draw[arrow] (x1.east) to[out=-20, in=180] (mul.west);
    \draw[arrow] (mul) -- (mm);
    \draw[arrow] (y1.east) to[out=0, in=180] (mm.west);
    \draw[arrow] (mm) -- (z1);
\end{scope}

% Original graph tiled
\begin{scope}[yshift=-5.1cm, local bounding box=partb]
    \fill[gray!15, rounded corners] (-0.7,1.4) rectangle (15,-4);

    \node[input] (eps) at (0, 0.7) {$\epsilon$};
    \node[input] (x1) at (0, -0.3) {$X$};
    \node[input] (y1) at (0, -3.2) {$W$};

    \node[recrmsop, align=center] (sq) at (2.5, 0) {[$n_0$, $n_1$, $b_0^{sq}$, $b_1^{sq}$] \\
   square $[t_0^{sq}, t_1^{sq}]$};
    \node[recrmsop, align=center] (mean) at (5.8, 0) {[$n_0$, $b_0^{m}$] \\
   mean $[t_0^{m}, 1]$};
    \node[recrmsop, align=center] (add) at (9, 0) {[$n_0$, $n_1$, $b_0^{a}$, $b_1^{a}$] \\
   add $[t_0^{a}, t_1^{a}]$};
    \node[recrmsop, align=center] (rsqrt) at (12.2, 0) {[$n_0$, $n_1$, $b_0^{r}$, $b_1^{r}$] \\
   rsqrt $[t_0^{r}, t_1^{r}]$};

    \node[recmulop, align=center] (mul) at (11.2, -1.6) {[$n_0$, $n_1$, $b_0^{ml}$, $b_1^{ml}$] \\
   multiply $[t_0^{ml}, t_1^{ml}]$};
    \node[recmatmulop, align=center] (mm) at (12, -3.2) {[$n_0$, $n_1$, $b_0^{m_0}$, $b_1^{m_0}$] \\
    matmul $[t_0^{m_0}, t_1^{m_0}]$};
    \node[output] (z1) at (14.4, -3.2) {$Z$};

    \node[text=black!60!black,
    align=left, font=\fontsize{15}{12}\selectfont] at (10, 1.7) {Perform tiling};
    \draw[->, line width=2pt] (8,2.2) -- (8,1);

    \draw[arrow] (x1.east) to[out=0, in=180] (sq.west);
    \draw[arrow] (sq) -- (mean);
    \draw[arrow] (mean) -- (add);
    \draw[arrow] (eps.east) to[out=10, in=160] (add.north west);
    \draw[arrow] (add) -- (rsqrt);
    \draw[arrow] (rsqrt.south) to[out=270, in=60] (mul.north);
    \draw[arrow] (x1.east) to[out=-30, in=180] (mul.west);
     \draw[arrow] (mul.south) to[out=270, in=150] (mm.north);
    \draw[arrow] (y1.east) to[out=0, in=180] (mm.west);
    \draw[arrow] (mm) -- (z1);
\end{scope}

\pgfresetboundingbox
\useasboundingbox (-0.7,1.6) rectangle (15,-9.4);

\end{tikzpicture}
  \vspace{-.2in}
  \subcaption{Original $\nu$Graph.}
  \label{fig:tiled-rmsnorm-a}
\end{minipage}
\hfill
\begin{minipage}[t]{0.45\textwidth}
  \centering
  \begin{tikzpicture}[
    scale=0.42,
    transform shape,
    input/.style={rectangle, draw=purple!70!black, fill=purple!30, minimum width=0.8cm, minimum height=0.7cm, font=\Large},
    output/.style={rectangle, draw=olive!70!black, fill=yellow!40, minimum width=0.8cm, minimum height=0.7cm, font=\Large},
    rmsop/.style={ellipse, draw=cyan!60!black, fill=cyan!40, minimum width=1.6cm, minimum height=0.9cm, font=\Large},
    mulop/.style={ellipse, draw=green!60!black, fill=green!35, minimum width=1.6cm, minimum height=0.9cm, font=\Large},
    matmulop/.style={ellipse, draw=red!40!orange, fill=red!25!orange!30, minimum width=1.6cm, minimum height=0.9cm, font=\Large},
    recrmsop/.style={rectangle, draw=cyan!60!black, fill=cyan!40, minimum width=1.6cm, minimum height=0.9cm, font=\Large},
    recmulop/.style={rectangle, draw=green!60!black, fill=green!35, minimum width=1.6cm, minimum height=0.9cm, font=\Large},
    recmatmulop/.style={rectangle, draw=red!40!orange, fill=red!25!orange!30, minimum width=1.6cm, minimum height=0.9cm, font=\Large},
    arrow/.style={->, >=stealth},
]

% Transformed graph (multiply propagated past matmul)
\begin{scope}[local bounding box=partc]
    \fill[gray!15, rounded corners] (-0.7,1.4) rectangle (14.8,-3);

    \node[input] (eps3) at (0, 0.7) {$\epsilon$};
    \node[input] (x3) at (0, -0.3) {$X$};
    \node[input] (y3) at (0, -2.3) {$W$};

    \node[rmsop] (sq3) at (2, 0) {square};
    \node[rmsop] (mean3) at (4.5, 0) {mean};
    \node[rmsop] (add3) at (7, 0) {add};
    \node[rmsop] (rsqrt3) at (9.5, 0) {rsqrt};

    \node[matmulop] (mm3) at (6.5, -2.4) {matmul};

    \node[mulop] (mul3) at (12, -1.2) {multiply};
    \node[output] (z3) at (14, -1.2) {$Z$};

    \draw[arrow] (x3.east) to[out=0, in=180] (sq3.west);
    \draw[arrow] (sq3) -- (mean3);
    \draw[arrow] (mean3) -- (add3);
    \draw[arrow] (eps3.east) to[out=10, in=160] (add3.north west);
    \draw[arrow] (add3) -- (rsqrt3);
    \draw[arrow] (rsqrt3.east) to[out=0, in=135] (mul3.north west);
    \draw[arrow] (x3.east) to[out=-60, in=120] (mm3.north);
    \draw[arrow] (y3.east) to[out=0, in=180] (mm3.west);
    \draw[arrow] (mm3.east) to[out=0, in=-135] (mul3.south west);
    \draw[arrow] (mul3) -- (z3);
\end{scope}

% Transformed graph tiled
\begin{scope}[yshift=-5.1cm, local bounding box=partd]
    \fill[gray!15, rounded corners] (-0.7,1.4) rectangle (15,-4);

    \node[input] (eps) at (0, 0.7) {$\epsilon$};
    \node[input] (x1) at (0, -0.3) {$X$};
    \node[input] (y1) at (0, -3.2) {$W$};

    \node[recrmsop, align=center] (sq) at (2.5, 0) {[$n_0$, $n_1$, $b_0^{sq}$, $b_1^{sq}$] \\
   square $[t_0^{sq}, t_1^{sq}]$};
    \node[recrmsop, align=center] (mean) at (5.8, 0) {[$n_0$, $b_0^{m}$] \\
   mean $[t_0^{m}, 1]$};
    \node[recrmsop, align=center] (add) at (9, 0) {[$n_0$, $n_1$, $b_0^{a}$, $b_1^{a}$] \\
   add $[t_0^{a}, t_1^{a}]$};
    \node[recrmsop, align=center] (rsqrt) at (12.2, 0) {[$n_0$, $n_1$, $b_0^{r}$, $b_1^{r}$] \\
   rsqrt $[t_0^{r}, t_1^{r}]$};

    \node[recmulop, align=center] (mul) at (12.2, -1.8) {[$n_0$, $n_1$, $b_0^{ml}$, $b_1^{ml}$] \\
   multiply $[t_0^{ml}, t_1^{ml}]$};
    \node[recmatmulop, align=center] (mm) at (6.5, -3.2) {[$n_0$, $n_1$, $b_0^{m_0}$, $b_1^{m_0}$] \\
    matmul $[t_0^{m_0}, t_1^{m_0}]$};
    \node[output] (z1) at (14.4, -3.2) {$Z$};

    \node[text=black!60!black,
    align=left, font=\fontsize{15}{12}\selectfont] at (10, 1.7) {Perform tiling};
    \draw[->, line width=2pt] (8,2.2) -- (8,1);

    \draw[arrow] (x1.east) to[out=0, in=180] (sq.west);
    \draw[arrow] (sq) -- (mean);
    \draw[arrow] (mean) -- (add);
    \draw[arrow] (eps.east) to[out=10, in=160] (add.north west);
    \draw[arrow] (add) -- (rsqrt);
    \draw[arrow] (rsqrt.south) to[out=270, in=90] (mul.north);
    \draw[arrow] (x1.east) to[out=-60, in=120] (mm.north);
     \draw[arrow] (mul.south) to[out=270, in=150] (z1.west);
    \draw[arrow] (y1.east) to[out=0, in=180] (mm.west);
    \draw[arrow] (mm.east) to[out=0, in=180] (mul.west);
\end{scope}

\pgfresetboundingbox
\useasboundingbox (-0.7,1.6) rectangle (15,-9.4);

\end{tikzpicture}
  \vspace{-.2in}
  \subcaption{Transformed $\nu$Graph.}
  \label{fig:tiled-rmsnorm-b}
\end{minipage}
\caption{Tiling of $\nu$Graphs for RMSNorm+MatMul. Each subfigure shows the high-level graph (top) and its tiled representation (bottom). In (b), \texttt{multiply} is propagated past \texttt{matmul}, creating two parallel branches that execute on different compute engines (Tensor Engine for matmul, Vector/Scalar Engine for RMS) with independent tiling parameters.}
\label{fig:tiled-rmsnorm}
\vspace{-.3in}
\end{figure*}

\paragraph{Tiling and Operator Reordering.}
Figure~\ref{fig:tiled-rmsnorm} shows how tiling interacts with the algebraic transformations from Section~\ref{subsec:tensor-graph-mutate}.
In the original graph (a), operations execute sequentially: \texttt{square}, \texttt{mean}, \texttt{rsqrt}, \texttt{multiply}, then \texttt{matmul}.
Each operation has its own tiling parameters (e.g., $[n_0, n_1, b_0^{sq},  b_1^{sq}]$ for \texttt{square}).
In the transformed graph (b), propagating \texttt{multiply} past \texttt{matmul} creates two parallel branches: the RMS computation (blue) and the matrix multiplication (red).
Because these branches execute on different compute engines with different hardware constraints---the Tensor Engine for matmul and the Vector/Scalar Engine for RMS---they can have \textit{independent} tiling parameters, enabling each branch to be optimized separately for its respective engine and on-chip memory budget.

% Table moved from background.tex — reader needs it here for ISA synthesis context
% NKI API Reference: https://awsdocs-neuron.readthedocs-hosted.com/en/latest/nki/api/nki.isa.html

% NKI API Reference: https://awsdocs-neuron.readthedocs-hosted.com/en/latest/nki/api/  nki.isa.html
 \begin{table*}[t]
     \footnotesize
     \centering
     \caption{A subset of supported NKI APIs that are representative of Trainium’s ISA.}
     \label{tab:nki-desc}
\scalebox{0.8}{
     \begin{tabular}{@{}p{5.8cm}p{8cm}l@{}}
     \toprule
     \textbf{API} & \textbf{Description} & \textbf{Engine} \\
     \midrule
     \multicolumn{3}{@{}l}{\textit{Data Movement}} \\
     % Ref: https://awsdocs-neuron.readthedocs-hosted.com/en/latest/nki/api/generated/  nki.isa.dma_copy.html
    \quad\texttt{dma\_copy(dst, src)} & Copy data between HBM and SBUF & DMA \\
     % Ref: https://awsdocs-neuron.readthedocs-hosted.com/en/latest/nki/api/generated/  nki.isa.tensor_copy.html
    \quad\texttt{tensor\_copy(src, engine)} & Create a copy of tile within on-chip     SRAMs (SBUF/PSUM) & Scalar/Vector/Gpsimd \\
     % Ref: https://awsdocs-neuron.readthedocs-hosted.com/en/latest/nki/api/generated/  nki.isa.dma_transpose.html
     \quad\texttt{dma\_transpose(src, axes)} & Transpose tile via DMA (supports 2D/3D/4D) & DMA \\
     % Ref: https://awsdocs-neuron.readthedocs-hosted.com/en/latest/nki/api/generated/  nki.isa.nc_transpose.html
     \quad\texttt{nc\_transpose(data, engine)} & 2D transpose between partition and     free axes of tile & Vector/Tensor \\
     \midrule
 \multicolumn{3}{@{}l}{\textit{Compute}} \\
     % Ref: https://awsdocs-neuron.readthedocs-hosted.com/en/latest/nki/api/generated/  nki.isa.nc_matmul.html
       \quad\texttt{nc\_matmul(stationary, moving)} & Matrix multiplication:              $\mathbf{stationary}^\top  \mathbf{moving}$ & Tensor \\
     % Ref: https://awsdocs-neuron.readthedocs-hosted.com/en/latest/nki/api/generated/  nki.isa.activation.html
     \quad\texttt{activation(op, data, bias, scale)} & Apply activation function with   optional scale/bias & Scalar \\
     % Ref: https://awsdocs-neuron.readthedocs-hosted.com/en/latest/nki/api/generated/  nki.isa.reciprocal.html
     \quad\texttt{reciprocal(data)} & Elementwise reciprocal of tile ($1.0/x$) &        Vector \\
     % Ref: https://awsdocs-neuron.readthedocs-hosted.com/en/latest/nki/api/generated/  nki.isa.tensor_tensor.html
     \quad\texttt{tensor\_tensor(data1, data2, op, engine)} & Elementwise binary        operation between two tiles & Vector/Gpsimd \\
     % Ref: https://awsdocs-neuron.readthedocs-hosted.com/en/latest/nki/api/generated/  nki.isa.tensor_scalar.html
     \quad\makecell[l]{\texttt{tensor\_scalar(data, op0, operand0,}\\\quad\texttt{op1,  operand1, engine)}} & \makecell[l]{Elementwise ops between \texttt{data} (a tile) and  \texttt{operand0}, \texttt{operand1}\\(scalars/vectors broadcast along free dim):  \\ \texttt{(data <op0> operand0) <op1> operand1}} & Vector/Scalar/Gpsimd \\
     % Ref: https://awsdocs-neuron.readthedocs-hosted.com/en/latest/nki/api/generated/  nki.isa.scalar_tensor_tensor.html
     \quad\makecell[l]{\texttt{scalar\_tensor\_tensor(data, op0,                        }\\\quad\texttt{operand0, op1, operand1)}} & \makecell[l]{Elementwise ops between      \texttt{data} (a tile), \texttt{operand0} (scalar/vector\\broadcast along free dim),   and \texttt{operand1} (a tile): \\  \texttt{(data <op0> operand0) <op1> operand1}} &       Vector \\     \midrule
     \multicolumn{3}{@{}l}{\textit{Reduction}} \\
     % Ref: https://awsdocs-neuron.readthedocs-hosted.com/en/latest/nki/api/generated/  nki.isa.tensor_reduce.html
     \quad\texttt{tensor\_reduce(op, data, axis)} & Reduction along free axes of tile & Vector \\
     % Ref: https://awsdocs-neuron.readthedocs-hosted.com/en/latest/nki/api/generated/  nki.isa.tensor_partition_reduce.html
     \quad\texttt{tensor\_partition\_reduce(op, data)} & Reduction across partitions of tile & Gpsimd \\
     % Ref: https://awsdocs-neuron.readthedocs-hosted.com/en/latest/nki/api/generated/  nki.isa.tensor_scalar_reduce.html
     \quad\makecell[l]{\texttt{tensor\_scalar\_reduce(data, op0,                        }\\\quad\texttt{operand0, reduce\_op)}} & \makecell[l]{Perform reduction               (\texttt{reduce\_op}) along free dim of \\ \texttt{(data <op0> operand0)}} & Vector \\

     % Ref: https://awsdocs-neuron.readthedocs-hosted.com/en/latest/nki/api/generated/  nki.isa.max8.html
     %\quad\texttt{max8(src)} & Find 8 largest values in each partition of tile          (descending order) & Scalar \\

     %\quad\\texttt{tensor\_tensor\_scan(data0, data1, \\ initial, op0, op1)} & Scan operation of two input tiles & Vector \\

     \quad\makecell[l]{\texttt{tensor\_tensor\_scan(data0, data1,                        }\\\quad\texttt{initial, op0, op1)}} & \makecell[l]{Scan operation of two input tiles} & Vector \\
     \bottomrule
     \end{tabular}
} %scalebox
\vspace{-.1in}
 \end{table*}

\subsection{Tensor ISA Synthesizer}
\label{subsec:synthesizer}

The Tensor ISA Synthesizer lowers each 2-D tensor operation in the Tiled $\nu$Graph to equivalent sequences of target ISA instructions using program synthesis.

\paragraph{ISA Semantics.}
\pname{} uses two levels of semantics specifications.
\textit{Tensor operator semantics} (Section~\ref{subsubsec:tensor-ops-sema}, Figure~\ref{fig:op-sema}) describe high-level operations such as \texttt{matmul} and \texttt{reduce\_sum}, and are provided by programmers writing tensor programs.
\textit{ISA instruction semantics} describe hardware-specific instructions such as Trainium's \texttt{nc\_matmul} and \texttt{nc\_transpose}, and are provided once per target by compiler developers with knowledge of the hardware.
Both use the same \texttt{@semantics} interface (preconditions, access maps, granular semantics), but ISA semantics additionally include \textit{hardware constraints} that specify valid tile dimension ranges for each instruction.
This separation means tensor operator semantics can be shared across targets, while ISA semantics are specified per target architecture.

Figure~\ref{fig:sema} shows the ISA semantics for Trainium's \texttt{nc\_matmul}, which computes $\mathbf{a}^\top  \mathbf{b}$.
The hardware constraints (\texttt{K <= 128}, \texttt{M <= 128}, \texttt{N <= 512}) reflect the Tensor Engine's tile dimension limits on Trainium.
These constraints are imposed on the symbolic tiling parameters from Section~\ref{subsec:tiling}, ensuring that synthesized code respects hardware limitations when emitting code in Section~\ref{subsec:code-emitter}.

\begin{wrapfigure}{r}{0.5\textwidth}
%\vspace{-5pt}
\begin{lstlisting}[style=pythonstyle, mathescape=true]
@semantics()
def nc_matmul (a : SymTensor, b : SymTensor):
    # Shape preconditions
    precondition (a.rank() == 2)
    precondition (b.rank() == 2)
    precondition (a.shape[0] == b.shape[0])
    K,M,N = a.shape[0], a.shape[1], b.shape[1]
    # Hardware constraints
    precondition (K <= 128)
    precondition (M <= 128)
    precondition (N <= 512)

    sema = OpSema (name="nc_matmul")
    sema.operands = [a, b]
    # Access maps
    i0 = SymAxis ("i0", range=[0, M])
    i1 = SymAxis ("i1", range=[0, K], is_contracting=True)
    i2 = SymAxis ("i2", range=[0, N])
    # Computation: a.T @ b
    sema.semantics = mul(a[i1, i0], b[i1, i2]).sum(axis=i1)
    sema.output_shape = [M, N]
    return sema
\end{lstlisting}

\caption{ISA semantics of Trainium's \texttt{nc\_matmul} instruction. Unlike the tensor operator semantics for \texttt{matmul} (Figure~\ref{fig:op-sema}), this specification includes hardware constraints and reflects the transposed access pattern (\texttt{a[i1, i0]}).}
\label{fig:sema}
\vspace{-15pt}
\end{wrapfigure}

\paragraph{Synthesis Algorithm.}
\pname{} uses a two-phase sketch-driven synthesis approach (Algorithm~\ref{alg:synthesizer}).
In Phase~1, \pname{} builds an instruction pool $\mathcal{P}$ by filtering ISA instructions to those sharing constituent operations with the tensor operation $T$ being lowered (lines 1--8).
For example, if $T$'s granular semantics (Section~\ref{subsubsec:tensor-ops-sema}) include  \texttt{add} and \texttt{multiply} operations, only ISA instructions that perform at least one of these operations are added to $\mathcal{P}$, reducing the search space. 
Instructions performing data layout transformations such as \texttt{transpose} are also included. 
The algorithm first retrieves the input operands $\mathcal{I}$ of $T$ (line 1)---for example, $\mathcal{I} = \{x, y\}$ for $\texttt{matmul}(x, y)$.
For each ISA instruction added to $\mathcal{P}$, \pname{} generates variants with its operand slots filled by either inputs from $\mathcal{I}$ or holes ($\square$) to be resolved during composition in Phase~2 (lines 4--8).
%
%When adding an instruction to $\mathcal{P}$, \pname{} places $\square$ (holes) or inputs to $T$ for its operands.
%\akchanged{, i.e., if $T$ comprises \texttt{add} and \texttt{multiply} operations, a target instruction is only added to $\mathcal{P}$ if it performs at least one of  these operations, thereby reducing the search space by not adding target operations that do not perform these operations. Operations performing data layouts transformations such as \texttt{transpose} are also added to $\mathcal{P}$. When adding an instruction to $\mathcal{P}$, \pname{} places $\square$ (holes) or inputs to $T$ for its operands.}
%
In Phase~2, \pname{} recursively fills holes in sketches with instructions from the pool (lines 9--19): starting from a sketch with a single hole, it pops candidates from a worklist, fills each hole with instructions from $\mathcal{P}$, and checks complete candidates for provable equivalence with $T$ using $\textsf{CheckEquivalence}$ (Section~\ref{sec:verification}).

\begin{algorithm}[t]
 \scriptsize
 \caption{\pname{}'s Sketch-driven Synthesis Algorithm}
 \label{alg:synthesizer}
 \begin{algorithmic}[1]
 \Require Tensor operation $T$, depth limit $d$
 \State $\mathcal{I} \gets \textsf{GetOperands}(T)$; \, $\mathcal{P}    \gets \{\mathcal{I}\}$ \Comment{Phase 1: build instruction pool}       \ForAll{$op$ in ISA instructions}
   \If{$\textsf{SharesConstituents}(op, T)$}
     \State $\mathcal{P} \gets \mathcal{P} \cup \{(op\ \square,\        \square)\}$ \Comment{$\square$ = hole}
     \ForAll{$in_1 \in \mathcal{I}$}
       \State $\mathcal{P} \gets \mathcal{P} \cup \{(op\ in_1,\         \square),\ (op\ \square,\ in_1)\}$
       \ForAll{$in_2 \in \mathcal{I}$}
         \State $\mathcal{P} \gets \mathcal{P} \cup \{(op\ in_1,\       in_2)\}$
       \EndFor
     \EndFor
   \EndIf
 \EndFor
 \State $\mathcal{W} \gets \{(\textbf{return}\ \square)\}$; \,          $\mathcal{C} \gets \varnothing$ \Comment{Phase 2: compose and check}  \While{$\mathcal{W} \neq \varnothing$}
   \State $s \gets \mathcal{W}.pop()$
   \If{$s$ has no holes}
     \If{$|s| \leq d$ \textbf{and} $\textsf{CheckEquivalence}(T, s)$}
       \State $\mathcal{C} \gets \mathcal{C} \cup \{s\}$                \Comment{valid candidate}
     \EndIf
    \Else
     \ForAll{hole $\square$ in $s$}
       \ForAll{$inst \in \mathcal{P}$}
         \State $\mathcal{W} \gets \mathcal{W} \cup \{s[\square         \mapsto inst]\}$ \Comment{fill hole}
       \EndFor
     \EndFor
   \EndIf
 \EndWhile
 \State \Return $\mathcal{C}$
 \end{algorithmic}
 \end{algorithm}

For example, when lowering \texttt{MatMul(x, y)}, the synthesizer      explores compositions such as:
\begin{equation*}
\operatorname{MatMul}(x, y)
\overset{?}{=}
\operatorname{nc\_matmul}\!\big(\operatorname{nc\_transpose}(x),\      y\big)
\end{equation*}
Since \texttt{nc\_matmul(a,b)} computes $\mathbf{a}^\top  \mathbf{b}$, the transpose is required to match standard matrix multiplication semantics.
On Trainium, an instruction depth limit of 3 is sufficient to synthesize all supported tensor operations, because the ISA is coarse-grained: each high-level tensor operation lowers to at most 3 ISA instructions.

\paragraph{Synthesis Example.}
Figure~\ref{fig:isa-graph} illustrates the synthesis process on the    Gated MLP kernel.
In part (a), the tiled $\nu$Graph contains high-level operations like \texttt{matmul}, \texttt{silu}, and \texttt{multiply}.
Part (b) shows the ISA $\nu$Graph after synthesis: notably, each \texttt{matmul} is lowered to \texttt{transpose} followed by \texttt{nc\_matmul}, because \texttt{nc\_matmul} expects its first operand to be transposed (as specified in the ISA semantics of Figure~\ref{fig:sema}).
The annotations show valid tile dimension ranges extracted from hardware constraints (e.g., \texttt{[1-128, 1-128]} for \texttt{nc\_matmul}).

\paragraph{DMA Operations and Engine Selection.}
After synthesizing compute instructions, \pname{} generates DMA copy instructions to transfer data between HBM and on-chip SRAM.
Additionally, since Trainium allows users to select which compute engine executes each instruction, \pname{} represents all valid engine variants in the $\nu$Graph, deferring the final selection to empirical evaluation.

\subsection{ISA Fusion Mutator}
\label{subsec:isa-egraph-mutator}

The ISA Fusion Mutator performs a second round of algebraic transformations (as described in Section~\ref{subsec:tensor-graph-mutate}), but this time at the ISA instruction level to expose ISA and operator (loop nest) fusion.
This second round is necessary because ISA synthesis (Section~\ref{subsec:synthesizer}) may introduce new operations that were not present in the original computation graph, and the new operations create transformation opportunities that the Graph Mutator (Section~\ref{subsec:tensor-graph-mutate}) could not have discovered. 

For example, in the Gated MLP kernel (Figure~\ref{fig:isa-graph} (a)), lowering \texttt{matmul} introduces a \texttt{transpose} operation since it is required by \texttt{nc\_matmul}'s semantics (Figure~\ref{fig:isa-graph} (b)).
%
%In this ISA $\nu$Graph, the \texttt{transpose} separates the element-wise operations (\texttt{activation}, \texttt{tensor\_tensor}) from the subsequent \texttt{nc\_matmul}, preventing fusion. 
\pname{} explores valid algebraic transformations on the newly generated ISA $\nu$Graph.
It discovers that the last \texttt{transpose} operation can be moved past the \texttt{activation} and \texttt{tensor\_tensor} instructions (Figure~\ref{fig:isa-graph} (c)), since both are element-wise instructions that commute with \texttt{transpose}.
%\akchanged{\pname explores what algebraic transformations on the newly generated ISA $\nu$Graph are valid, and deems moving the last \texttt{transpose} operation past the \texttt{activation} and \texttt{tensor\_tensor} instructions as shown in Figure~\ref{fig:isa-graph} (c) as a valid transformation because both are element-wise instructions that commute with a \texttt{transpose}.}
%
%The ISA Fusion Mutator propagates the element-wise operations past \texttt{transpose} (c), which is valid because element-wise operations commute with layout transformations.
%
%\akchanged{Since no more algebraic transformations lead to discovery of variants, \pname{} looks for fusion opportunities on adjacent nodes of the $\nu$Graphs. This leads to discovery of two similar fusion opportunities: the two \texttt{nc\_matmul} operations and succeeding \texttt{transpose} operations are deemed fusible, and are later fused in Figure~\ref{fig:isa-graph} (d).}
Once there are no further algebraic transformations that yield new variants, \pname{} looks for fusion opportunities on adjacent nodes of the $\nu$Graph. 
It identifies two fusion opportunities: the two \texttt{nc\_matmul} operations and their succeeding \texttt{transpose} operations are fusible, producing the graph in Figure~\ref{fig:isa-graph} (d).

%now have matching block dimensions, enabling them to be fused into a single loop nest that keeps intermediate results in on-chip memory.

The mutator performs two types of fusion:

\begin{wrapfigure}{r}{0.5\textwidth}
\centering
\vspace{-15pt}
\scalebox{0.8}{
\raisebox{0pt}[\height][0pt]{\input{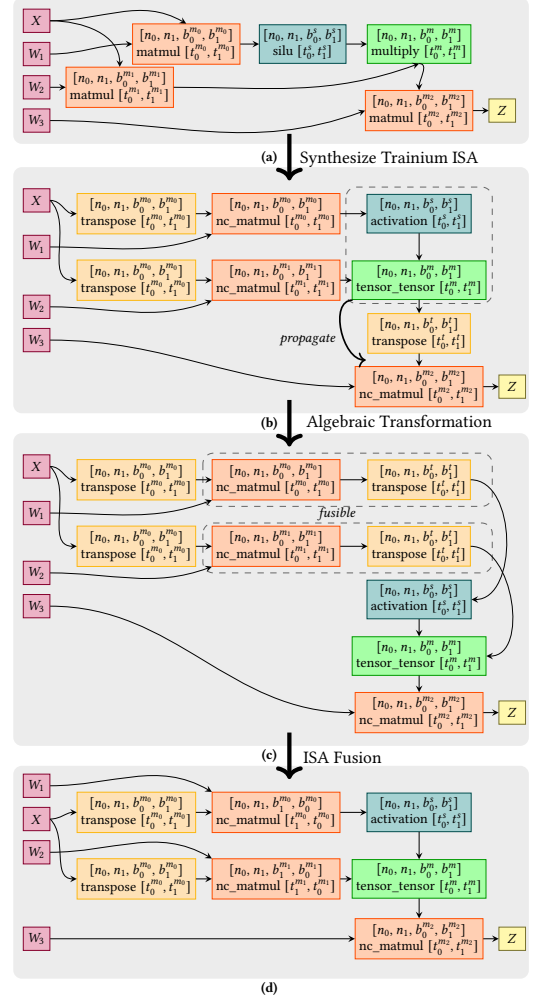}}
}
\caption{(a) Tiled $\nu$Graph for Gated MLP; (b) ISA $\nu$Graph after synthesis; (c) Mutated ISA $\nu$Graph after moving \texttt{activation} and \texttt{tensor\_tensor} past \texttt{transpose}, resulting in discovery of fusible operations; 
(d) ISA $\nu$Graph after fusion---the operands of \texttt{nc\_matmul} and their tile and block dimensions are swapped. Instruction-specific tile size constraints and engine choices are omitted for brevity.
}
\vspace{-25pt}
\label{fig:isa-graph}
\end{wrapfigure}

\paragraph{Instruction Fusion.} The mutator performs peephole optimizations by replacing two or more instructions in ISA $\nu$Graphs with a single, semantically-equivalent instruction.
This is achieved using the same synthesis algorithm as Section~\ref{subsec:synthesizer} (Algorithm~\ref{alg:synthesizer}), but with depth 1 since the goal is to fuse two instructions into one.
In Figure~\ref{fig:isa-graph}, \texttt{nc\_matmul} and the succeeding \texttt{transpose} operations are fused into a new \texttt{nc\_matmul} but with swapped operands---\pname{} discovers via synthesis that $\operatorname{transpose} (\operatorname{nc\_matmul}(X.t, W)) \rightarrow \operatorname{nc\_matmul}(W, X.t)$.
This effectively eliminates an intermediate data layout transformation. Superoptimizers such as Mirage~\cite{wu2025mirage} do not support such a rewrite.
%This transformation effectively eliminates an intermediate data layout transformation. Superoptimizers such as Mirage \cite{wu2025mirage} do not support such a rewrite rule.
%The mutator applies peephole optimizations to replace sequences of ISA instructions with fused variants.

Other examples include fusing \texttt{activation} and \texttt{tensor\_reduce} to \texttt{activation\_reduce}, and \texttt{tensor\_scalar} and \texttt{tensor\_reduce} to \texttt{tensor\_scalar\_reduce}. 
This pass also fuses \texttt{dma\_copy} with \texttt{dma\_transpose} to allow on-the-fly \texttt{transpose} when reading from or writing to HBM.

%For instance, on Trainium, a \texttt{dma\_copy} followed by \texttt{nc\_transpose} can be replaced with \texttt{dma\_transpose}, performing the layout transformation on-the-fly during data transfer.

\paragraph{Operator/Loop Fusion.} This is more similar to the traditional fusion optimization: consecutive nodes of the $\nu$Graphs with the same strip dimensions, across which no reduction occurs, are deemed fusible along those dimensions.  (Reductions such as \texttt{mean} in RMSNorm~\cite{rmsnorm} require processing an entire tensor row before normalization, preventing fusion across that dimension.)
This improves data locality and decreases memory traffic between HBM and SBUF. 
\pname{} also fuses nodes of independent subgraphs to enable deep fusion of loop nests, leveraging compute and data movement parallelism across different engines.

%share compatible tiling dimensions, their loop nests can be merged so that intermediate results remain in on-chip memory rather than spilling to HBM.
%
%Since \pname{} explores multiple tiling configurations (Section~\ref{subsec:tiling}), matching dimensions between consecutive operations arise naturally from the search; the mutator identifies and exploits these matches. 

% The mutator explores all valid reorderings and fusion choices, retaining them in the $\nu$Graph for empirical evaluation.

\subsection{NKI Code Emitter}
\label{subsec:code-emitter}

The NKI Code Emitter is the final stage of \pname{}'s synthesis pipeline.
It traverses the mutated ISA $\nu$Graph to extract concrete NKI programs, where each program represents a specific combination of algebraic rewrite, instruction selection, tiling configuration, and fusion decisions.
The symbolic block sizes from Section~\ref{subsec:tiling} are instantiated with concrete values, generating multiple candidates per $\nu$Graph.

Each emitted NKI program is then passed to the Neuron compiler~\cite{neuron-compiler}, which handles on-chip buffer allocation (assigning SBUF and PSUM addresses to intermediate tensors) and produces the final executable.
Because \pname{} does not directly control on-chip SRAM allocation, some candidates that satisfy \pname{}'s symbolic tile constraints may still be rejected by the Neuron compiler---for example, when the cumulative size of live tiles in a block exceeds the physical SBUF capacity.
This is why not all emitted programs are compilable (Section~\ref{subsec:comp-times}).

\textit{Empirical execution and random testing.} For the candidates that are successfully compiled to binaries by the Neuron compiler, \pname{} executes them on the target hardware, measures latency, and validates numerical correctness against a reference implementation on random inputs.
We deem the outputs numerically equivalent if they satisfy, elementwise, $\lvert \mathrm{emitted\_code}-\mathrm{reference}\rvert \le \texttt{atol}+\texttt{rtol}\,\lvert \mathrm{reference}\rvert$ with $\texttt{rtol}=10^{-4}$ (relative error) and $\texttt{atol}=10^{-4}$ (absolute error) on FP32 data type. %\zsw{Q: what datatype did we use in these kernels, I assume all FP32? Need to be clear on this if we give the atol rtol.}
\pname{} then returns the best-performing NKI program along with its configuration (tile sizes, instruction choices, fusion decisions).
This empirical selection ensures that the final kernel is optimized for the specific input shapes and target hardware.

\section{Correctness of Transformations}
\label{sec:verification}

Every transformation that \pname{} performs---whether reordering operators in the computation graph (Section~\ref{subsec:tensor-graph-mutate}), synthesizing ISA instruction sequences (Section~\ref{subsec:synthesizer}) or fusing instructions (Section~\ref{subsec:isa-egraph-mutator})---must preserve semantic equivalence.
This section describes how \pname{} leverages SMT-based reasoning to provably check that each transformation is correct.

\subsection{Equivalence Checking Methodology}
\label{subsec:equiv-checking}

\pname{}'s equivalence checking adapts the methodology introduced by TensorRight~\cite{arora2025tensorright}---originally designed to verify XLA rewrite rules---to provably check transformations from operator propagation (Section~\ref{subsec:tensor-graph-mutate}) and ISA synthesis (Section~\ref{subsec:synthesizer}).
%\pname{}'s equivalence checking builds on the methodology introduced by TensorRight~\cite{arora2025tensorright}, which was originally designed to prove correctness of rewrite rules for XLA.
%
%\pname{} adapts this methodology to provably check whether transformations produced by the operator propagation (Section~\ref{subsec:tensor-graph-mutate}) and ISA synthesis (Section~\ref{subsec:synthesizer}) stages preserve semantic equivalence.

\paragraph{Encoding.}
To provably check that two tensor expressions are equivalent, \pname{} encodes each operation's computation as SMT constraints and checks two conditions:
\begin{enumerate}
\item \textbf{\textit{Shape Condition:}} The symbolic shapes of output tensors on both sides must be equivalent.
\item \textbf{\textit{Computation Condition:}} The computation performed on each element must be equivalent.
\end{enumerate}

Tensors are encoded as \textit{uninterpreted functions} mapping indices to values (e.g., a rank-2 tensor $X$ becomes $X: \mathbb{Z} \times \mathbb{Z} \to \mathbb{R}$), and dimension sizes are symbolic integers.
This allows reasoning over tensors of \textit{arbitrary size} with known rank---the proof holds for any concrete dimension sizes.
Table~\ref{tab:z3-encoding} summarizes how different categories of operations are encoded.

\begin{table*}[t]
\footnotesize
\centering
\caption{SMT encoding of a subset of tensor operations, following TensorRight's methodology~\cite{arora2025tensorright}. Element-wise and layout operations are checked directly by the SMT solver. Reductions use arithmetic axioms. Nonlinear functions are treated as uninterpreted and conservatively rejected when swapped.}
\label{tab:z3-encoding}
\scalebox{0.82}{
\begin{tabular}{@{}llll@{}}
\toprule
\textbf{Category} & \textbf{Operations} & \textbf{SMT Encoding} & \textbf{Axioms} \\
\midrule
Element-wise & mul, div, add, sub & $\forall i,j{:}\; out(i,j) = a(i,j) \circ b(i,j)$ & None \\
Unary pointwise & sqrt, relu & $\forall i,j{:}\; out(i,j) = f(a(i,j))$ & None \\
Reduction & reduce\_sum & $P(i,0){=}0;\;\; P(i,k{+}1){=}P(i,k){+}a(i,k);\;\; out(i){=}P(i,N)$ & Arithmetic axioms \\
Matmul & matmul & $P(i,j,0){=}0;\;\; P(i,j,k{+}1){=}P(i,j,k){+}a(i,k){\cdot}b(k,j);\;\; out(i,j){=}P(i,j,K)$ & Arithmetic axioms \\
Nonlinear & exp, sigmoid & Uninterpreted function (minimal properties) & Conservative rejection \\
Layout & transpose & $\forall i,j{:}\; out(i,j) = a(j,i)$ & None \\
\bottomrule
\end{tabular}
}
\vspace{-.12in}
\end{table*}

Table~\ref{tab:nki-ops} shows the granular operations for key tensor operations and NKI ISA instructions used in the equivalence checking.

For element-wise and layout operations, the encoding is straightforward and the solver can determine equivalence directly.
For reductions and matrix multiplications, which involve summation over a contraction dimension, the methodology provides arithmetic axioms---such as the distributivity of multiplication over summation ($\operatorname{mul}(\sum f(x),\,v) = \sum \operatorname{mul}(f(x),\,v)$ when $v$ is invariant over the summation)---that enable the solver to reason about equivalence of reduction expressions.
Nonlinear functions (e.g., \texttt{exp}, \texttt{sigmoid}) are treated as uninterpreted functions: the solver knows only minimal properties (e.g., $\exp(x) > 0$) and cannot reason about their algebraic behavior, so swaps involving them are conservatively rejected.

\paragraph{Checking procedure.}
Given two adjacent operators $op_1$ and $op_2$ in the computation graph, \pname{} checks whether they can be swapped:
\begin{enumerate}
\item Construct the swapped graph by exchanging $op_1$ and $op_2$.
\item Encode both the original and swapped graphs as SMT constraints using the encodings in Table~\ref{tab:z3-encoding}.
\item Check the Shape Condition: prove that output shapes are equal under all valid input shapes.
\item Check the Computation Condition: prove that for all valid indices, the output values are equal---i.e., $\forall \vec{i}{:}\; out_{\text{orig}}(\vec{i}) = out_{\text{swap}}(\vec{i})$.
\item If the solver proves both conditions, the swap is accepted. If the solver returns \texttt{unknown} (e.g., timeout), the swap is conservatively rejected.
\end{enumerate}

\begin{definition}[Swappable successive tensor operations]
\label{def:swappable}
Let $OP_1$ and $OP_2$ be two successive operations in a computation graph $G$.
$OP_1$ and $OP_2$ are deemed \textit{swappable} if the solver proves that applying $OP_1$ then $OP_2$ produces the same output as applying $OP_2$ then $OP_1$:
\[
  \llbracket OP_2 \rrbracket\bigl(\llbracket OP_1 \rrbracket(\vec{x})\bigr)
  \;\;\equiv\;\;
  \llbracket OP_1 \rrbracket\bigl(\llbracket OP_2 \rrbracket(\vec{x})\bigr)
\]
where $\llbracket \cdot \rrbracket$ denotes the semantics of an operation and $\vec{x}$ represents the input tensors.
\end{definition}

\begin{wraptable}{r}{0.55\textwidth}
\vspace{-0.5\baselineskip} 
\footnotesize
\centering
\caption{Granular operations for a subset of tensor and ISA operations used for equivalence checking.}
\label{tab:nki-ops}
\scalebox{0.82}{
\begin{tabular}{lc}
\toprule
{\textbf{Operator}} & {\textbf{Granular Operation}} \\
\midrule
matmul(x, y) & $sum(mul(x_{i,k}, y_{k,j}), dim=k)$ \\
transpose(x) & $x_{i,k} \to x_{k,i}$ \\
relu(x) & $max(0, x_{i,j})$ \\
silu(x) & $mul(x_{i,j}, sigmoid(x_{i,j}))$ \\
\midrule
nc\_matmul(x, y) & $sum(mul(x_{k,i}, y_{k,j}), dim=k)$ \\
nc\_transpose(x, engine) & $x_{i,k} \to x_{k,i}$ \\
tensor\_tensor(x1, x2, op, engine) & $op(x_{i,j}, y_{i,j})$ \\
tensor\_reduce(op, x, axis) & $op(x_{i,j, red\_dim=j})$ \\
tensor\_partition\_reduce(op, x) & $op(x_{i,j, red\_dim=i})$ \\
activation(op, x) & $op(x_{i,j})$ \\
tensor\_scalar(x, op1, y, op2, z, engine) & $op2(op1(x_{i,j}, y_{i,j}), z_{i,j})$ \\
scalar\_tensor\_tensor(x, op1, y, op2, z) & $op2(op1(x_{i,j}, y_{i,j}), z_{i,j})$ \\
tensor\_scalar\_reduce(x, op1, y, op2, z) & $op2(op1(x_{i,j}, y_{i,j}), z_{i,j})$ \\
\bottomrule
\end{tabular}
}
\vspace{-0.5\baselineskip}
\end{wraptable}

\paragraph{Examples.}

\textit{Legal swap: multiply past matmul.}
In the RMSNorm+MatMul kernel (Section~\ref{sec:motivation}), \pname{} checks whether $\operatorname{MatMul}(\operatorname{Multiply}(X, rsqrt), W) \overset{?}{=} \\ \operatorname{Multiply}(\operatorname{MatMul}(X, W), rsqrt)$.
For the Shape Condition: both sides produce shape $[M, N]$ when $X$ is $[M, K]$, $rsqrt$ is $[M, 1]$, and $W$ is $[K, N]$.
For the Computation Condition: the LHS computes $\sum_k (X_{i,k} \cdot rsqrt_i) \cdot W_{k,j}$ and the RHS computes $(\sum_k X_{i,k} \cdot W_{k,j}) \cdot rsqrt_i$.
Since $rsqrt_i$ is invariant over $k$, the distributivity axiom enables the solver to confirm equivalence. The swap is accepted.

\textit{Illegal swap: silu past multiply.}
In the Gated MLP kernel (Section~\ref{subsec:tensor-graph-mutate}), \pname{} checks whether $\operatorname{Multiply}(\operatorname{Silu}(X), G) \overset{?}{=} \operatorname{Silu}(\operatorname{Multiply}(X, G))$.
The Shape Condition is satisfied (both element-wise, same output shape).
For the Computation Condition: the LHS computes $X_{i,j} \cdot \sigma(X_{i,j}) \cdot G_{i,j}$ and the RHS computes $X_{i,j} \cdot G_{i,j} \cdot \sigma(X_{i,j} \cdot G_{i,j})$, where $\sigma$ is the sigmoid function.
Since $\sigma$ is an uninterpreted function, the solver cannot prove these are equal. The swap is rejected.

\textit{ISA synthesis: matmul via nc\_matmul.}
During ISA synthesis (Section~\ref{subsec:synthesizer}), \pname{} checks whether $\operatorname{MatMul}(x, y) \overset{?}{=} \operatorname{nc\_matmul}(\operatorname{nc\_transpose}(x), y)$.
Since $\operatorname{nc\_matmul}(a, b)$ computes $a^T \cdot b$ (Table~\ref{tab:nki-ops}), transposing $x$ first recovers standard multiplication.
The solver confirms both conditions, accepting the synthesized instruction sequence.

\subsection{Guarantees and Limitations}
\label{subsec:verify-limits}

\pname{}'s equivalence checking guarantees semantic equivalence over all valid inputs for tensors of arbitrary size with known rank.
The checking is conservative: if the solver cannot prove equivalence, the transformation is rejected, ensuring no incorrect transformations are applied.
This provides stronger correctness guarantees than probabilistic equivalence testing (as used by Mirage~\cite{wu2025mirage}), which tests on finite inputs and provides statistical but not provable guarantees.

The approach has two limitations. 
%
%\zsw{Q: how about dynamic control flow and dynamic shapes? This version of NKI does not support but is there any fundamental limitation of the technique?} \ak{We can support dynamic shapes easily because all the verification framework operates on unbounded tensors.} \zsw{Sounds good! maybe should mention it explicitly?}
%
First, the equivalence checking operates over real-valued arithmetic and does not account for floating-point precision---the generated kernels may exhibit small numerical differences due to rounding, which is standard in tensor program optimization.
We use the solver's real theory because it is significantly faster than its floating-point theory: as an example, swapping \texttt{multiply} and \texttt{matmul} in the RMSNorm+MatMul kernel alone (Figure~\ref{fig:rmsnorm-matmul-comp-graph}) takes 247.75\,s with floating-point theory and 0.15\,s with real theory (1650$\times$ slowdown). 
Floating-point theory is prohibitively slow for program synthesis and verification \cite{z3issue, izycheva2020counterexample, li2014symbolic, trentin2019optimization, conchon2012built}.
Instead, \pname{} uses the real theory for algebraic transformations and ISA synthesis, and validates numerical equivalence via testing on random floating-point inputs (Section~\ref{subsec:code-emitter}).
%\akchanged{(\pname{} uses testing on random inputs to ensure numerical equivalence as discussed in Section~\ref{subsec:comp-times}.)}
%
Second, nonlinear functions are treated as uninterpreted, which means the solver may conservatively reject valid swaps involving these functions. 

Dynamic shapes are naturally supported by the equivalence checking, since it operates on symbolic dimensions over unbounded tensors and the proofs hold for any concrete dimension sizes.

\section{Evaluation}
\label{sec:eval}

\subsection{Implementation and Experimental Setup}
\label{subsec:exp-setup}

\pname{} is implemented in approximately 16,500 lines of Python 3.10~\cite{van2007python}.
The equivalence checking engine (Section~\ref{sec:verification}) reimplements TensorRight's~\cite{arora2025tensorright} methodology in Python, using Z3 4.12.6~\cite{de2008z3} as the SMT solver.
\pname{} takes NumPy-like tensor programs as input and emits NKI programs~\cite{aws_neuron_nki_index}---the tile-level kernel language for Trainium---which are then processed by the Neuron compiler~\cite{neuron-compiler}.

\textbf{Benchmarks.} Our benchmark suite includes 11 individual operators---MatMul, RMSNorm, Cumsum, Softmax, RoPE, Tensor Multiply, Max Pooling, ReLU, SiLU, SwiGLU, and GLU---and 9 multi-operator kernels: RMSNorm+MatMul, Matmul+Add+RMSNorm, Softmax+MatMul, Transpose+MatMul, QKV Projection, Group Query Attention (GQA), SiLU Gated MLP, SwiGLU Gated MLP, and Matmul+Matmul+SiLU+Mul.
These cover the core computational kernels in modern LLM architectures such as LLaMA-3~\cite{grattafiori2024llama} and Falcon~\cite{almazrouei2023falcon}, and significantly overlap with the benchmarks evaluated in Mirage~\cite{wu2025mirage}.
For each benchmark, we sweep across a range of tensor dimension sizes; since performance depends only on tensor shapes, we repeat each experiment 100 times (and 20 times for warmup) using random inputs and report the average run time.

\noindent\textbf{Baselines.} We compare \pname against the following baselines:

\begin{itemize}

     \item \textbf{Amazon's Neuron compiler.} The Neuron compiler~\cite{neuron-compiler} (version \texttt{2.21.18209.0+043b1bf7}) is the proprietary commercial compiler for Trainium that performs graph-level optimizations and uses hand-tuned heuristics for code generation.
     As the primary production compiler for Trainium, it serves as a key baseline.

     \item \textbf{Hand-optimized NKI kernels.} Amazon provides hand-optimized NKI kernels in the \texttt{nki-samples} repository~\cite{aws-neuron_nki-samples_2025}.
     Note that hand-optimized NKI kernels are available for only a small subset of operators.

     \item \textbf{Search-based compiler.} Mirage~\cite{wu2025mirage} has been recently extended to target Trainium for a handful of kernels (RMSNorm+MatMul and Partial Gated MLP).
     The NKI code generation extension\footnote{\url{https://github.com/mirage-project/mirage/blob/d669c42174db27bea0455a24e20768b01b7d52e4/docs/transpiler/nki_transpiler.md}} was obtained by contacting the authors directly.
     It uses hand-implemented templates for individual operators such as MatMul and element-wise operators, and composes them based on the kernel specification.

\end{itemize}

All compilation paths to Trainium go through the Neuron compiler, but enter at different levels: \pname{} takes NumPy-like programs and emits NKI; Mirage takes custom kernel specifications implemented in C++ and also emits NKI; the Neuron compiler baseline takes tensor programs written using NumPy~\cite{harris2020array} and handles all transformations and optimizations---tiling, operator fusion, instruction selection---internally with its own heuristics; and hand-optimized NKI kernels are written directly in NKI.
All emitted NKI programs are processed by the same Neuron compiler backend.
All baselines use their default configurations.
For the hand-optimized NKI kernels, we use commit \texttt{7af4e4d} of the \texttt{nki-samples} repository~\cite{aws-neuron_nki-samples_2025} and the NKI language version is \texttt{2.21.18209.0}.

\noindent\textbf{Hardware Platform.} We run all experiments on an Amazon EC2 \texttt{trn1.32xlarge} instance equipped with 16 Trainium chips with 32 Neuron cores and 128 cores of Intel Xeon Platinum 8375C processors (2.90\,GHz, 55\,MB L3 cache) with hyperthreading disabled.
We perform our experiments on only one Neuron core to evaluate the effectiveness of \pname's optimization and code generation on a single core.

We structure our evaluation around four questions:
\begin{itemize}[leftmargin=*]
     \item \textbf{RQ1:} How does \pname{} perform against existing baselines on individual tensor operators? Although these kernels pass through the full \pname{} pipeline, the performance gains are primarily due to tiling (Section~\ref{subsec:tiling}) and ISA instruction selection (Section~\ref{subsec:synthesizer}): individual operators have limited opportunities for algebraic transformations or inter-operator fusion.

     \item \textbf{RQ2:} How does \pname{} perform on multi-operator kernels, where inter-operator algebraic transformations (Section~\ref{subsec:tensor-graph-mutate}) and loop fusion across operators (Section~\ref{subsec:isa-egraph-mutator}) provide additional optimization opportunities?

     \item \textbf{RQ3:} What is the contribution of each optimization (such as algebraic transformations, fusion)?

     \item \textbf{RQ4:} What is the compilation overhead of \pname{}, and how does it scale with kernel complexity?
\end{itemize}

Not all baselines are available for every benchmark: the Neuron compiler does not support kernels with inter-loop dependencies (e.g., Cumsum); Mirage's NKI extension only supports MatMul, RMSNorm+MatMul, Softmax+MatMul, and Matmul+Matmul+SiLU+Mul; and hand-optimized NKI kernels are only available for Cumsum and RoPE.

\begin{table}[t]
\centering
\caption{\pname{} speedups on individual tensor operators against the Neuron Compiler and hand-implemented NKI across different input dimension sizes (M$\times$N; 1K=1024, 2K=2048, 4K=4096, 8K=8192, 16K=16384). ``--'' indicates the NKI kernel not supporting that dimension size. Shading: \colorbox{blue!15}{$<$1.5x}, \colorbox{blue!30}{1.5--2.5x}, \colorbox{blue!50}{\textcolor{white}{$>$2.5x}}.}

\label{tab:single-op-eval}
\setlength{\tabcolsep}{1.2pt}
\renewcommand{\arraystretch}{1.00}
\scriptsize
\begin{tabular}{l*{25}{c}c}
\toprule
\textbf{Benchmark} &
\rotatebox{90}{\textbf{1Kx1K}} &
\rotatebox{90}{\textbf{1Kx2K}} &
\rotatebox{90}{\textbf{1Kx4K}} &
\rotatebox{90}{\textbf{1Kx8K}} &
\rotatebox{90}{\textbf{1Kx16K}} &
\rotatebox{90}{\textbf{2Kx1K}} &
\rotatebox{90}{\textbf{2Kx2K}} &
\rotatebox{90}{\textbf{2Kx4K}} &
\rotatebox{90}{\textbf{2Kx8K}} &
\rotatebox{90}{\textbf{2Kx16K}} &
\rotatebox{90}{\textbf{4Kx1K}} &
\rotatebox{90}{\textbf{4Kx2K}} &
\rotatebox{90}{\textbf{4Kx4K}} &
\rotatebox{90}{\textbf{4Kx8K}} &
\rotatebox{90}{\textbf{4Kx16K}} &
\rotatebox{90}{\textbf{8Kx1K}} &
\rotatebox{90}{\textbf{8Kx2K}} &
\rotatebox{90}{\textbf{8Kx4K}} &
\rotatebox{90}{\textbf{8Kx8K}} &
\rotatebox{90}{\textbf{8Kx16K}} &
\rotatebox{90}{\textbf{16Kx1K}} &
\rotatebox{90}{\textbf{16Kx2K}} &
\rotatebox{90}{\textbf{16Kx4K}} &
\rotatebox{90}{\textbf{16Kx8K}} &
\rotatebox{90}{\textbf{16Kx16K}} &
\rotatebox{90}{\textbf{Geomean}} \\
\midrule
\multicolumn{27}{l}{\textbf{vs Neuron Compiler}} \\
\midrule
Multiply & \hm{1.0} & \hm{1.0} & \hm{1.0} & \hm{1.0} & \hm{1.0} & \hm{1.0} & \hm{1.0} & \hm{1.0} & \hm{1.0} & \hm{1.5} & \hm{1.0} & \hm{1.0} & \hm{1.0} & \hm{1.5} & \hm{1.9} & \hm{1.0} & \hm{1.0} & \hm{1.5} & \hm{1.9} & \hm{2.9} & \hm{1.0} & \hm{1.5} & \hm{1.8} & \hm{2.8} & \hm{2.8} & \hm{1.23} \\
RMSNorm  & \hm{1.0} & \hm{1.0} & \hm{1.0} & \hm{1.0} & \hm{1.4} & \hm{1.0} & \hm{1.0} & \hm{1.0} & \hm{1.4} & \hm{2.2} & \hm{1.1} & \hm{1.0} & \hm{1.3} & \hm{2.3} & \hm{3.0} & \hm{1.0} & \hm{1.4} & \hm{2.1} & \hm{3.3} & \hm{3.0} & \hm{1.4} & \hm{2.1} & \hm{2.8} & \hm{3.3} & \hm{3.0} & \hm{1.62} \\
ReLU     & \hm{1.0} & \hm{1.0} & \hm{1.0} & \hm{1.2} & \hm{1.6} & \hm{1.0} & \hm{1.0} & \hm{1.2} & \hm{1.6} & \hm{2.2} & \hm{1.0} & \hm{1.0} & \hm{1.6} & \hm{2.3} & \hm{2.9} & \hm{1.0} & \hm{1.6} & \hm{2.2} & \hm{3.0} & \hm{3.1} & \hm{1.6} & \hm{2.1} & \hm{2.9} & \hm{3.1} & \hm{3.1} & \hm{1.70} \\
SiLU     & \hm{1.0} & \hm{1.1} & \hm{1.3} & \hm{1.2} & \hm{1.6} & \hm{1.1} & \hm{1.3} & \hm{1.2} & \hm{1.6} & \hm{2.4} & \hm{1.3} & \hm{1.2} & \hm{1.6} & \hm{2.4} & \hm{3.4} & \hm{1.2} & \hm{1.6} & \hm{2.4} & \hm{3.5} & \hm{3.6} & \hm{1.6} & \hm{2.4} & \hm{3.4} & \hm{3.6} & \hm{3.7} & \hm{1.90} \\
GLU      & \hm{1.0} & \hm{1.1} & \hm{1.3} & \hm{1.2} & \hm{1.4} & \hm{1.1} & \hm{1.2} & \hm{1.2} & \hm{1.4} & \hm{2.0} & \hm{1.3} & \hm{1.2} & \hm{1.6} & \hm{2.0} & \hm{3.4} & \hm{1.2} & \hm{1.6} & \hm{2.0} & \hm{3.1} & \hm{3.2} & \hm{1.3} & \hm{2.0} & \hm{3.1} & \hm{3.3} & \hm{3.3} & \hm{1.77} \\
SwiGLU   & \hm{1.0} & \hm{1.3} & \hm{1.1} & \hm{1.1} & \hm{1.5} & \hm{1.3} & \hm{1.0} & \hm{1.1} & \hm{1.5} & \hm{2.1} & \hm{1.1} & \hm{1.1} & \hm{1.6} & \hm{2.2} & \hm{2.9} & \hm{1.1} & \hm{1.5} & \hm{2.3} & \hm{3.0} & \hm{2.9} & \hm{1.5} & \hm{2.1} & \hm{3.1} & \hm{3.0} & \hm{3.0} & \hm{1.67} \\
Softmax  & \hm{1.0} & \hm{1.0} & \hm{1.0} & \hm{1.0} & \hm{1.3} & \hm{1.0} & \hm{1.0} & \hm{1.0} & \hm{1.2} & \hm{2.1} & \hm{1.0} & \hm{1.0} & \hm{1.3} & \hm{2.0} & \hm{2.9} & \hm{1.0} & \hm{1.3} & \hm{1.8} & \hm{2.7} & \hm{2.9} & \hm{1.2} & \hm{2.0} & \hm{2.3} & \hm{2.7} & \hm{3.0} & \hm{1.52} \\
MaxPool  & \hm{1.0} & \hm{1.0} & \hm{1.0} & \hm{1.1} & \hm{1.2} & \hm{1.0} & \hm{1.0} & \hm{1.0} & \hm{1.2} & \hm{1.3} & \hm{1.0} & \hm{1.0} & \hm{1.0} & \hm{1.5} & \hm{1.7} & \hm{1.0} & \hm{1.0} & \hm{1.5} & \hm{1.5} & \hm{2.2} & \hm{1.0} & \hm{1.5} & \hm{1.4} & \hm{2.2} & \hm{2.1} & \hm{1.23} \\
\midrule
\multicolumn{27}{l}{\textbf{vs Hand-implemented NKI}} \\
\midrule
Cumsum &
-- & \hm{1.00} & \hm{1.35} & \hm{1.00} & \hm{1.08} &
-- & \hm{1.35} & \hm{1.00} & \hm{1.03} & \hm{1.07} &
-- & \hm{1.00} & \hm{1.02} & \hm{1.08} & \hm{1.08} &
-- & \hm{1.01} & \hm{1.03} & \hm{1.08} & \hm{1.10} &
-- & \hm{1.01} & \hm{1.04} & \hm{1.10} & \hm{1.08} & \hm{1.07} \\
RoPE &
-- & \hm{1.00} & \hm{1.20} & \hm{1.00} & \hm{1.07} &
-- & \hm{1.10} & \hm{1.00} & \hm{1.03} & \hm{1.06} &
-- & \hm{1.00} & \hm{1.02} & \hm{1.07} & \hm{1.08} &
-- & \hm{1.01} & \hm{1.01} & \hm{1.08} & \hm{1.11} &
-- & \hm{1.01} & \hm{1.06} & \hm{1.11} & \hm{1.08} & \hm{1.06} \\
\bottomrule
\end{tabular}
\vspace{-15pt}
\end{table}

\subsection{RQ1: Individual Operators}
\label{subsec:rq1}

\begin{table}[t]
\centering
\caption{Speedups of \pname{} over the Neuron Compiler and Mirage on multi-operator kernels. MM=MatMul, Proj=Projection, GQA=Group Query Attention; 2K/4K/8K=2048/4096/8192. Mirage results shown only for kernels it could synthesize. Shading: \colorbox{blue!15}{1.0--1.1x},
  \colorbox{blue!30}{1.1--1.2x},
  \colorbox{blue!50}{\textcolor{white}{$>$1.2x}}.}
\label{tab:multi-op-eval}
\setlength{\tabcolsep}{1.2pt}
\renewcommand{\arraystretch}{1.05}
\scriptsize
\scalebox{0.87}{
\begin{tabular}{l*{27}{c}c}
\toprule
\textbf{Benchmark} &
\rotatebox{90}{\textbf{2Kx2Kx2K}} &
\rotatebox{90}{\textbf{2Kx2Kx4K}} &
\rotatebox{90}{\textbf{2Kx2Kx8K}} &
\rotatebox{90}{\textbf{2Kx4Kx2K}} &
\rotatebox{90}{\textbf{2Kx4Kx4K}} &
\rotatebox{90}{\textbf{2Kx4Kx8K}} &
\rotatebox{90}{\textbf{2Kx8Kx2K}} &
\rotatebox{90}{\textbf{2Kx8Kx4K}} &
\rotatebox{90}{\textbf{2Kx8Kx8K}} &
\rotatebox{90}{\textbf{4Kx2Kx2K}} &
\rotatebox{90}{\textbf{4Kx2Kx4K}} &
\rotatebox{90}{\textbf{4Kx2Kx8K}} &
\rotatebox{90}{\textbf{4Kx4Kx2K}} &
\rotatebox{90}{\textbf{4Kx4Kx4K}} &
\rotatebox{90}{\textbf{4Kx4Kx8K}} &
\rotatebox{90}{\textbf{4Kx8Kx2K}} &
\rotatebox{90}{\textbf{4Kx8Kx4K}} &
\rotatebox{90}{\textbf{4Kx8Kx8K}} &
\rotatebox{90}{\textbf{8Kx2Kx2K}} &
\rotatebox{90}{\textbf{8Kx2Kx4K}} &
\rotatebox{90}{\textbf{8Kx2Kx8K}} &
\rotatebox{90}{\textbf{8Kx4Kx2K}} &
\rotatebox{90}{\textbf{8Kx4Kx4K}} &
\rotatebox{90}{\textbf{8Kx4Kx8K}} &
\rotatebox{90}{\textbf{8Kx8Kx2K}} &
\rotatebox{90}{\textbf{8Kx8Kx4K}} &
\rotatebox{90}{\textbf{8Kx8Kx8K}} &
\rotatebox{90}{\textbf{Geomean}} \\
\midrule
\multicolumn{29}{l}{\textbf{vs Neuron Compiler}} \\
\midrule
Transpose+MM &
\hmc{1.01} & \hmc{1.01} & \hmc{1.00} &
\hmc{1.04} & \hmc{1.06} & \hmc{1.01} &
\hmc{1.05} & \hmc{1.05} & \hmc{1.06} &
\hmc{1.06} & \hmc{1.01} & \hmc{1.06} &
\hmc{1.01} & \hmc{1.06} & \hmc{1.03} &
\hmc{1.09} & \hmc{1.11} & \hmc{1.15} &
\hmc{1.22} & \hmc{1.00} & \hmc{1.03} &
\hmc{1.00} & \hmc{1.06} & \hmc{1.02} &
\hmc{5.61} & \hmc{1.14} & \hmc{19.01} &
\hmc{1.32} \\
RMSNorm+MM &
\hmc{1.40} & \hmc{1.00} & \hmc{1.22} &
\hmc{1.00} & \hmc{1.28} & \hmc{1.00} &
\hmc{1.03} & \hmc{1.12} & \hmc{1.06} &
\hmc{1.47} & \hmc{1.00} & \hmc{1.25} &
\hmc{1.00} & \hmc{1.27} & \hmc{1.00} &
\hmc{1.03} & \hmc{1.12} & \hmc{1.07} &
\hmc{1.42} & \hmc{1.00} & \hmc{1.28} &
\hmc{1.00} & \hmc{1.23} & \hmc{1.00} &
\hmc{1.03} & \hmc{1.13} & \hmc{1.07} &
\hmc{1.11} \\
MM+Add+RMSNorm &
\hmc{1.51} & \hmc{1.41} & \hmc{1.45} &
\hmc{1.32} & \hmc{1.34} & \hmc{1.23} &
\hmc{1.13} & \hmc{1.03} & \hmc{1.02} &
\hmc{1.05} & \hmc{1.00} & \hmc{1.05} &
\hmc{1.00} & \hmc{1.09} & \hmc{1.00} &
\hmc{1.03} & \hmc{1.02} & \hmc{1.14} &
\hmc{1.08} & \hmc{1.05} & \hmc{1.00} &
\hmc{1.00} & \hmc{2.25} & \hmc{1.00} &
\hmc{1.02} & \hmc{1.06} & \hmc{1.03} &
\hmc{1.11} \\
Softmax+MM &
\hmc{1.47} & \hmc{1.00} & \hmc{1.19} &
\hmc{1.13} & \hmc{1.38} & \hmc{1.04} &
\hmc{1.06} & \hmc{1.14} & \hmc{1.09} &
\hmc{1.47} & \hmc{1.00} & \hmc{1.17} &
\hmc{1.12} & \hmc{1.36} & \hmc{1.04} &
\hmc{1.06} & \hmc{1.15} & \hmc{1.07} &
\hmc{1.64} & \hmc{1.00} & \hmc{1.17} &
\hmc{1.13} & \hmc{1.37} & \hmc{1.04} &
\hmc{1.06} & \hmc{1.17} & \hmc{1.08} &
\hmc{1.14} \\
MM+MM+SiLU+Mul &
\hmc{1.22} & \hmc{1.12} & \hmc{1.14} &
\hmc{1.11} & \hmc{1.18} & \hmc{1.18} &
\hmc{1.28} & \hmc{1.17} & \hmc{1.12} &
\hmc{1.30} & \hmc{1.11} & \hmc{1.12} &
\hmc{1.11} & \hmc{1.16} & \hmc{1.14} &
\hmc{1.29} & \hmc{1.16} & \hmc{1.15} &
\hmc{1.18} & \hmc{1.13} & \hmc{1.09} &
\hmc{1.05} & \hmc{1.13} & \hmc{1.06} &
\hmc{1.28} & \hmc{1.14} & \hmc{1.13} &
\hmc{1.15} \\
QKV Projection &
\hmc{2.35} & \hmc{1.00} & \hmc{2.12} &
\hmc{1.00} & \hmc{1.06} & \hmc{1.02} &
\hmc{1.08} & \hmc{1.02} & \hmc{1.03} &
\hmc{2.37} & \hmc{1.00} & \hmc{2.12} &
\hmc{1.04} & \hmc{1.09} & \hmc{1.01} &
\hmc{1.04} & \hmc{1.03} & \hmc{1.16} &
\hmc{2.39} & \hmc{1.95} & \hmc{2.14} &
\hmc{1.04} & \hmc{1.50} & \hmc{1.00} &
\hmc{1.02} & \hmc{1.07} & \hmc{1.02} &
\hmc{1.24} \\
GQA &
\hmc{1.47} & \hmc{1.00} & \hmc{1.19} &
\hmc{1.13} & \hmc{1.38} & \hmc{1.04} &
\hmc{1.06} & \hmc{1.14} & \hmc{1.09} &
\hmc{1.47} & \hmc{1.00} & \hmc{1.17} &
\hmc{1.12} & \hmc{1.36} & \hmc{1.04} &
\hmc{1.06} & \hmc{1.15} & \hmc{1.07} &
\hmc{1.64} & \hmc{1.00} & \hmc{1.17} &
\hmc{1.13} & \hmc{1.37} & \hmc{1.04} &
\hmc{1.06} & \hmc{1.17} & \hmc{1.08} &
\hmc{1.14} \\
SiLU MLP &
\hmc{1.07} & \hmc{1.00} & \hmc{1.50} &
\hmc{1.02} & \hmc{1.08} & \hmc{1.04} &
\hmc{1.10} & \hmc{1.09} & \hmc{1.06} &
\hmc{1.17} & \hmc{1.00} & \hmc{1.52} &
\hmc{1.02} & \hmc{1.08} & \hmc{1.02} &
\hmc{1.07} & \hmc{1.09} & \hmc{1.07} &
\hmc{1.13} & \hmc{1.00} & \hmc{1.54} &
\hmc{1.04} & \hmc{1.09} & \hmc{1.00} &
\hmc{1.07} & \hmc{1.09} & \hmc{1.06} &
\hmc{1.08} \\
SwiGLU MLP &
\hmc{1.07} & \hmc{1.00} & \hmc{1.30} &
\hmc{1.02} & \hmc{1.08} & \hmc{1.04} &
\hmc{1.10} & \hmc{1.09} & \hmc{1.06} &
\hmc{1.17} & \hmc{1.00} & \hmc{1.52} &
\hmc{1.02} & \hmc{1.08} & \hmc{1.02} &
\hmc{1.07} & \hmc{1.09} & \hmc{1.07} &
\hmc{1.13} & \hmc{1.00} & \hmc{1.54} &
\hmc{1.04} & \hmc{1.09} & \hmc{1.00} &
\hmc{1.07} & \hmc{1.09} & \hmc{1.06} &
\hmc{1.08} \\
\midrule
\multicolumn{29}{l}{\textbf{vs Mirage}} \\
\midrule
RMSNorm+MM &
\hmc{1.08} & \hmc{1.11} & \hmc{1.03} &
\hmc{1.15} & \hmc{1.11} & \hmc{1.05} &
\hmc{1.08} & \hmc{1.06} & \hmc{1.06} &
\hmc{1.11} & \hmc{1.11} & \hmc{1.11} &
\hmc{1.22} & \hmc{1.11} & \hmc{1.03} &
\hmc{1.07} & \hmc{1.06} & \hmc{1.06} &
\hmc{1.10} & \hmc{1.15} & \hmc{1.13} &
\hmc{1.22} & \hmc{1.13} & \hmc{1.08} &
\hmc{1.08} & \hmc{1.07} & \hmc{1.06} &
\hmc{1.10} \\
Softmax+MM &
\hmc{1.00} & \hmc{1.22} & \hmc{1.07} &
\hmc{1.13} & \hmc{1.00} & \hmc{1.21} &
\hmc{1.13} & \hmc{1.06} & \hmc{1.09} &
\hmc{1.01} & \hmc{1.16} & \hmc{1.06} &
\hmc{1.11} & \hmc{1.02} & \hmc{1.18} &
\hmc{1.10} & \hmc{1.05} & \hmc{1.08} &
\hmc{1.00} & \hmc{1.19} & \hmc{1.05} &
\hmc{1.10} & \hmc{1.01} & \hmc{1.16} &
\hmc{1.14} & \hmc{1.06} & \hmc{1.09} &
\hmc{1.06} \\
MM+MM+SiLU+Mul &
\hmc{1.11} & \hmc{1.01} & \hmc{1.04} &
\hmc{1.09} & \hmc{1.09} & \hmc{1.01} &
\hmc{1.16} & \hmc{1.15} & \hmc{1.11} &
\hmc{1.24} & \hmc{1.17} & \hmc{1.12} &
\hmc{1.21} & \hmc{1.09} & \hmc{1.18} &
\hmc{1.25} & \hmc{1.15} & \hmc{1.03} &
\hmc{1.12} & \hmc{1.10} & \hmc{1.12} &
\hmc{1.11} & \hmc{1.02} & \hmc{1.08} &
\hmc{1.27} & \hmc{1.08} & \hmc{1.11} &
\hmc{1.10} \\
\bottomrule
\end{tabular}
}
\vspace{-10pt}
\end{table}

\begin{figure*}[t]
   \centering
   \includegraphics[width=0.99\linewidth]{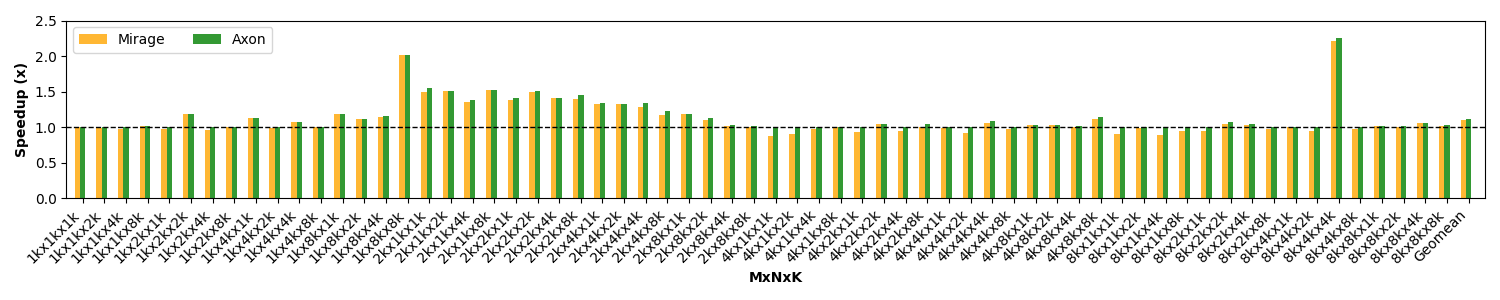}
   \vspace{-.1in}
   \caption{Speedups of \pname{} and Mirage on Matrix Multiplication relative to the Neuron Compiler across different input dimension sizes (M$\times$N$\times$K). MatMul has three dimensions, thus separate from Table~\ref{tab:single-op-eval}. %\chungha{can we change to 2K, 4K, 8K for x-axis and add caption?}
   }
   \label{fig:matmul-results}
\vspace{-.2in}   
\end{figure*}

Table~\ref{tab:single-op-eval} and Figure~\ref{fig:matmul-results} show the performance of \pname{} on individual tensor operators across a range of tensor dimension sizes.

\textbf{Against the Neuron Compiler.}
Across all operators compared against the Neuron compiler---Tensor Multiplication, RMSNorm, Softmax, ReLU, SiLU, GLU, SwiGLU, and Max Pooling (Table~\ref{tab:single-op-eval}, top)---\pname{} matches or slightly outperforms the Neuron compiler at smaller dimension sizes and achieves speedups of up to 3.7x (SiLU at 16384$\times$16384) at larger sizes, with geomean speedups ranging from 1.23x (Multiply, MaxPool) to 1.90x (SiLU).
The speedups at larger sizes are primarily because the code synthesized by \pname{} achieves better temporal locality for tiles of data in on-chip buffer and better DMA utilization.
The Neuron compiler uses fixed heuristics for tiling that become suboptimal as tensor dimensions grow, whereas \pname{} systematically explores the space of valid tiling configurations and selects the best-performing one empirically.
%\chungha{Make sure the last sentence is correct} \ak{this is correct.}

\textbf{Against Mirage.}
Mirage provides a hand-tuned NKI template only for MatMul among the individual operators.
On Matrix Multiplication (Figure~\ref{fig:matmul-results}), \pname{} achieves performance comparable to Mirage across all dimension sizes, with a geomean speedup of 2\% and a maximum speedup of 14\%.
%\chungha{what's the max and geomean value? We need the max and geomean}.
%
This result is notable because Mirage's MatMul template is manually crafted and parameterized for different tiling factors, while \pname{} discovers its tiling automatically.

\textbf{Against hand-optimized NKI kernels.}
For Cumsum and RoPE (Table~\ref{tab:single-op-eval}, bottom), hand-optimized NKI kernels are the only available baseline.
\pname{} matches the expert code on most configurations and outperforms them by up to 1.35x on certain configurations (geomean speedups of 1.07x and 1.06x respectively), demonstrating that \pname{}'s automated exploration can match and even surpass manually tuned code without requiring the programmer to have hardware expertise.

\subsection{RQ2: Multi-Operator Kernels}
\label{subsec:rq2}

Table~\ref{tab:multi-op-eval} shows the performance of \pname{} on multi-operator kernels.
Unlike the individual operator benchmarks, these kernels benefit from the combination of \pname{}'s optimization strategies---algebraic transformations, operator and ISA fusion, and effective tiling---to achieve speedups.

\textbf{Against the Neuron Compiler.}
\pname{} achieves speedups over the Neuron compiler across all multi-operator benchmarks (Table~\ref{tab:multi-op-eval}, top).
On these benchmarks, \pname{} achieves better performance by discovering independent subgraphs using algebraic transformations and then deeply fusing the loop nests, thereby maximizing data reuse across tensor operators and enabling hardware to better exploit pipeline parallelism; whereas the Neuron compiler fails to fuse computations across tensor operators, which often leads to costly spillage of data into HBM.
For RMSNorm+MatMul, the running example from Section~\ref{sec:motivation}, \pname{} achieves up to 1.47x speedup with a geomean of 1.11x.
The $\nu$Graph mutator discovers that the element-wise multiply can be reordered past the matrix multiplication (Figure~\ref{fig:rmsnorm-matmul-comp-graph}), enabling the RMS computation on the Vector/Scalar Engine to execute in parallel with the matrix multiplication on the Tensor Engine, while fusion keeps normalization intermediates in on-chip buffer.
MatMul+Add+RMSNorm achieves up to 2.25x (geomean 1.11x), and Matmul+Matmul+SiLU+Mul---which performs part of the Gated MLP computation---achieves up to 1.30x (geomean 1.15x).
The full Gated MLP layers (SiLU and SwiGLU activated) achieve geomean speedups of 1.08x.
Transpose+MatMul achieves a geomean speedup of 1.32x, with improvements growing as dimensions increase: at smaller dimensions \pname{} achieves modest speedups of 1--15\%, while at the largest dimensions the gains are 5.6x (8192$\times$8192$\times$2048) and 19x (8192$\times$8192$\times$8192).
We attribute these large speedups to \pname{}'s better tiling strategy and efficient DMA bandwidth utilization: at high dimensions, the gap between a suboptimal tiling configuration and an optimal one widens because poorly chosen tile sizes cause more frequent spillage from on-chip buffer to HBM and poor DMA utilization. 
\pname{}'s systematic exploration of tiling configurations finds tile block dimensions that keep data on-chip and improve DMA utilization, while a heuristic-based approach is more likely to miss these configurations as the search space grows.
QKV Projection achieves up to 2.39x (geomean 1.24x) and GQA achieves up to 1.64x (geomean 1.14x).

\textbf{Against Mirage.}
Mirage's NKI extension supports three multi-operator benchmarks (Table~\ref{tab:multi-op-eval}, bottom): RMSNorm+MatMul, Softmax+MatMul, and Matmul+Matmul+SiLU+Mul.
The remaining multi-operator benchmarks are not supported because Mirage's NKI code generation relies on hand-implemented templates that do not yet cover these operator compositions.
On the benchmarks Mirage does support, \pname{} achieves geomean speedups of 10\% on RMSNorm+MatMul, 6\% on Softmax+MatMul, and 10\% on Matmul+Matmul+SiLU+Mul.
Notably, Mirage's hand-crafted NKI templates sometimes produce code that is slower than the Neuron compiler baseline (e.g., on several dimension sizes for RMSNorm+MatMul in Table~\ref{tab:multi-op-eval}), whereas \pname{} consistently matches or outperforms the Neuron compiler through systematic exploration of algebraic transformations and fusion opportunities not supported by Mirage.

\subsection{RQ3: Ablation Study on Optimizations}
\label{subsec:ablation}

To understand the contribution of each optimization to performance, we selectively disable algebraic transformations and operator fusion, and measure the impact on four multi-operator benchmarks (Table~\ref{tab:ablation}).
We do not disable tiling because it is required for executing kernels with large tensors that do not fit in the on-chip buffers.
As discussed in Section~\ref{subsec:rq1}, the individual operator benchmarks (Table~\ref{tab:single-op-eval}) primarily demonstrate the impact of tiling, since single-operator kernels have limited opportunities for algebraic transformations or inter-operator fusion.

\textbf{Operator fusion is the most impactful optimization.}
Across all four benchmarks, disabling fusion causes the largest performance degradation, with geomean slowdowns of 35--50\%.
Without fusion, intermediate results between operators must be written to and read back from HBM, significantly increasing memory traffic.

\textbf{Algebraic transformations provide consistent gains.}
Disabling algebraic transformations leads to geomean slowdowns of 15--35\% across the benchmarks.
The impact is present across all four benchmarks, confirming that the graph mutator's operator reordering contributes to performance even when fusion is still enabled.

\textbf{The two optimizations are complementary.}
The combined effect of both optimizations (i.e., the full \pname pipeline) consistently outperforms either optimization alone, indicating that algebraic transformations and fusion target different sources of inefficiency: algebraic transformations expose parallelism and reduce unnecessary computation; fusion reduces memory traffic between operators.

\begin{table}[t]

\centering
\caption{Ablation study: performance relative to \pname{} with all optimizations enabled, measured across all 27 dimension sizes from Table~\ref{tab:multi-op-eval}. Values below 1.0x indicate slowdown. Per-dimension results are in Appendix~\ref{app:ablation-figures}.%\chungha{Please fill the numbers}
}
\label{tab:ablation}
\small
\scalebox{0.85}{
\begin{tabular}{l ccc ccc}
\toprule
& \multicolumn{3}{c}{\textbf{No Algebraic Transform}} & \multicolumn{3}{c}{\textbf{No Loop Fusion}} \\
\cmidrule(lr){2-4} \cmidrule(lr){5-7}
\textbf{Benchmark} & Min & Max & Geomean & Min & Max & Geomean \\
\midrule
RMSNorm+MatMul      & 0.50x & 0.90x & 0.65x & 0.33x & 0.80x & 0.53x \\
QKV Projection       & 0.39x & 0.98x & 0.85x & 0.24x & 0.90x & 0.65x \\
Group Query Attention & 0.50x & 0.98x & 0.77x & 0.34x & 0.96x & 0.60x \\
SiLU MLP            & 0.51x & 0.94x & 0.80x & 0.21x & 0.92x & 0.64x \\
\bottomrule
\end{tabular}
}
\vspace{-.1in}
\end{table}

\subsection{RQ4: Compilation Overhead}
\label{subsec:comp-times}

Table~\ref{tab:compile-times} reports the compilation overhead for all benchmarks.

\begin{itemize}
     \item \textbf{Phase~1 (Lowering and Emission):} The full \pname pipeline---graph mutation, tiling, ISA synthesis, ISA fusion mutation, and NKI code emission---transforms the input computation graph and emits multiple candidate NKI programs.

     \item \textbf{Phase~2 (Compilation and Execution):} Each emitted NKI program is compiled by the Neuron compiler backend and, if compilation succeeds, executed on Trainium. The best-performing kernel is selected.
\end{itemize}

Table~\ref{tab:compile-times} reports, for each benchmark, the number of graph-level operations (\textbf{\# Ops}), the Phase~1 time, the number of emitted and compilable NKI programs, the Phase~2 time, and the total time. 
Note that \textbf{\# Ops} counts operations after decomposition: a single high-level operator such as RMSNorm decomposes into multiple graph operations (square, mean, rsqrt, multiply), so \# Ops can exceed 1 even for individual operator benchmarks.
%Table~\ref{tab:compile-times} reports six quantities for each benchmark: the number of operations in the computation graph (\textbf{\# Ops})---note that a single high-level operator such as RMSNorm decomposes into multiple graph operations (e.g., square, mean, rsqrt, multiply), so \# Ops can exceed 1 even for individual operator benchmarks; the time spent in Phase~1; the number of NKI programs emitted by Phase~1 (\textbf{\# Emitted}); the number of those that successfully compile via the Neuron compiler backend (\textbf{\# Compilable}); the time spent in Phase~2 compiling and executing the compilable kernels; and the total time.
%
Not all emitted programs are compilable: as described in Section~\ref{subsec:code-emitter}, \pname{} does not directly control on-chip buffer allocation, so candidates that satisfy \pname{}'s symbolic tile constraints may still be rejected by the Neuron compiler when the cumulative size of live tiles in a block exceeds the physical SBUF capacity.

\textbf{Phase~1 is fast; Phase~2 dominates compilation time.}
Phase~1 completes in under 82 seconds even for the most complex kernels, as it operates symbolically on the computation graph.
Phase~2, which involves parallelized compilation and serialized execution of NKI kernels on hardware, accounts for over 97\% of total compilation time.
Its cost is proportional to the number of compilable NKI kernels: element-wise and reduction operators produce 456--1{,}126 compilable kernels and complete Phase~2 in 289--348 seconds; kernels involving matrix multiplication produce 5{,}000--18{,}982 compilable kernels and take 3{,}156--4{,}107 seconds; and complex LLM layers produce 18{,}450--19{,}934 compilable kernels and take 4{,}992--5{,}983 seconds.

\textbf{The search space is driven by the number of operators, input dimensions, and hardware constraints.}
In general, more operators (\textbf{\# Ops}) lead to more emitted candidates, since each operator introduces tiling configurations and instruction choices that compose multiplicatively (Section~\ref{subsec:nu-graphs}).
However, the number of input dimensions and hardware constraints also play a significant role.
MatMul, despite being a single operator, emits 14{,}554 candidates---far more than other individual operators (625--1{,}475)---because its three input dimensions (M$\times$K$\times$N) create a larger combinatorial space for block size exploration compared to the two dimensions of element-wise operators.
Conversely, Transpose+MatMul (2 operators) emits only 8{,}050 candidates, fewer than MatMul alone, because the ISA Fusion Mutator (Section~\ref{subsec:isa-egraph-mutator}) eliminates redundant \texttt{transpose} instructions: lowering \texttt{matmul} introduces a \texttt{transpose} (required by \texttt{nc\_matmul}'s semantics), which is fused with the pre-existing \texttt{transpose} operation in the input graph, reducing the number of emitted candidates.
%\akchanged{the variants of \texttt{transpose} instructions introduced from the synthesis of \texttt{nc\_matmul} after lowering of the \texttt{matmul} operation are fused with the pre-existing \texttt{transpose} operation and eliminated before code generation (the \texttt{transpose} is required by \texttt{nc\_matmul}'s semantics), thereby generating fewer NKI candidates}.
%
Among multi-operator kernels, the search space grows with complexity but is bounded by hardware constraints: GQA (6 operators) and Gated MLP (5--7 operators) approach an upper bound of approximately 29{,}500 emitted candidates.

\textbf{Not all emitted candidates are compilable.}
Across all benchmarks, 60--76\% of emitted kernels are compilable (e.g., GQA: 19,627 of 28,219); the remainder are rejected by the Neuron compiler due to hardware resource constraints such as insufficient SBUF capacity (Section~\ref{subsec:code-emitter}).
%
%\cschanged{
%This suggests pruning infeasible candidates before invoking the backend compiler---for example, by estimating the cumulative on-chip footprint of live tiles within a block---could further reduce compilation time.
%}
%As discussed in Section~\ref{subsec:code-emitter}, some candidates that pass \pname{}'s symbolic constraints are rejected by the Neuron compiler when live tiles exceed SBUF capacity.
%
%Across all benchmarks, 60--76\% of emitted kernels are compilable (e.g., GQA: 19{,}627 of 28{,}219), which saves Phase~2 time for the rejected candidates.
%
%This also presents an opportunity for \pname{} to prune infeasible candidates before invoking the backend compiler, further reducing compilation time (e.g., by estimating the cumulative on-chip footprint of live tiles within a block).

\textbf{The compilation overhead is a one-time cost.}
The total time ranges from approximately 5 minutes for simple operators to 1.7 hours for the most complex kernels (SwiGLU Gated MLP); since performance depends only on tensor shapes, the optimized kernel is reused for all subsequent executions.
This one-time search cost is comparable to that of other optimizing compilers that explore large search spaces:
Mirage~\cite{wu2025mirage} reports search times of up to 4 hours on GPUs for similar LLM kernels, and schedule-based compilers such as Halide~\cite{ragan2013halide}, AutoTVM~\cite{chen2018learning}, and Ansor~\cite{zheng2020ansor} take several hours to explore tiling and fusion choices.
In practice, kernels for commonly used tensor shapes can be compiled ahead of time and cached as artifacts for reuse across models---a standard practice at companies like Amazon to amortize compilation cost~\cite{mccarthy2025enhanced}.

Appendix~\ref{app:threats} discusses threats to validity, including end-to-end inference, generalizability beyond Trainium, and benchmark coverage.

%---including end-to-end inference, generalizability beyond Trainium, and benchmark coverage---

\begin{table}[h!]
\centering
\vspace{-.1in}
\caption{Compilation overhead of \pname{}. \textbf{\textit{Phase 1:}} lowering and emission of candidate programs (mutation, tiling, synthesis, fusion, emission); \textbf{\textit{Phase 2:}} compilation and execution on Trainium.}
\label{tab:compile-times}
\scalebox{0.75}{
\begin{tabular}{l r r r r r r}
\toprule
\textbf{Kernel} & \textbf{\# Ops} & \textbf{Phase 1 (s)} & \textbf{\# Emitted} & \textbf{\# Compilable} & \textbf{Phase 2 (s)} & \textbf{Total (s)} \\ \midrule
Tensor Multiply       & 1 & 8.2  & 625      & 456      & 288.5    & 296.7   \\
Cumsum                & 1 & 11.4 & 625      & 456      & 314.3    & 325.7   \\
RMSNorm               & 5 & 15.6 & 900      & 540      & 348.2    & 363.8   \\
Softmax               & 3 & 16.3 & 900      & 540      & 319.9    & 336.2   \\
ReLU                  & 1 & 9.6  & 625      & 456      & 290.5    & 300.1   \\
SiLU                  & 1 & 10.8 & 625      & 456      & 304.1    & 314.9   \\
GLU                   & 3 & 10.8 & 1,258    & 875      & 305.7    & 316.5   \\
SwiGLU                & 3 & 10.9 & 1,258    & 875      & 308.9    & 319.8   \\
MaxPool               & 1 & 22.7 & 1,475    & 1,126    & 324.5    & 347.2   \\
RoPE                  & 1 & 15.9 & 1,250    & 912      & 327.2    & 343.1   \\ 
MatMul                & 1 & 24.8 & 14,554   & 9,532    & 3,228.6  & 3,253.4 \\ \midrule
Transpose+MatMul      & 2 & 29.3 & 8,050    & 5,000    & 3,155.7  & 3,185.0 \\
RMSNorm+MatMul        & 6 & 37.1 & 17,500   & 12,750   & 3,246.4  & 3,283.5 \\
MatMul+Add+RMSNorm    & 7 & 39.4 & 19,500   & 14,758   & 3,253.7  & 3,293.1 \\
Softmax+MatMul        & 4 & 31.9 & 16,263   & 12,549   & 3,242.4  & 3,274.3 \\
MM+MM+SiLU+Mul        & 4 & 75.4 & 27,740   & 18,982   & 4,106.8  & 4,182.2 \\ \midrule
QKV Projection         & 3 & 68.2 & 26,750   & 18,450   & 5,878.9  & 5,947.1 \\
Group Query Attention  & 6 & 79.4 & 28,219   & 19,627   & 4,992.2  & 5,071.6 \\
SiLU Gated MLP         & 5 & 81.6 & 29,535   & 19,934   & 5,949.3  & 6,030.9 \\
SwiGLU Gated MLP       & 7 & 79.8 & 29,535   & 19,934   & 5,982.8  & 6,062.6 \\
\bottomrule
\end{tabular}
}
\vspace{-.1in}
\end{table}

\section{Related Work}
\label{sec:relwork}

\textbf{Algebraic transformation compilers.}
TASO~\cite{jia2019taso}, PET~\cite{wang2021pet}, Grappler~\cite{grappler}, and EinNet~\cite{zheng2023einnet} optimize tensor programs by rewriting computation graphs using algebraic equivalences: TASO enumerates user-provided rewrite rules with random testing, and PET adds partially equivalent rewrites with correction kernels.
Equality saturation-based compilers (e.g., Tensat~\cite{tensat}, Constable~\cite{vohra2025mind}) apply hand-written Egg~\cite{willsey2021egg} rewrite rules but provide no semantic correctness guarantees; overall, these methods rely on pattern-specific rules and do not automatically reason about operator semantics. The closest prior work to \pname{} is Mirage~\cite{wu2025mirage}, which uses probabilistic equivalence checking to jointly optimize algebraic and scheduling transformations on $\mu$Graphs and relies on hand-written NKI templates to constrain search for Trainium.
In contrast, \pname{} discovers transformations using operator propagation, synthesizes ISA instructions from specifications, and proves correctness with SMT over unbounded tensors, without hand-written rewrite rules or templates.

\textbf{Schedule-based compilers.}
Halide~\cite{ragan2013halide} introduced the idea of separating the algorithm from its schedule; TVM~\cite{chen2018tvm}, Ansor~\cite{zheng2020ansor}, and others~\cite{zheng2020flextensor, hagedorn2023graphene, feng2022tensorir} build on this idea to search for optimized schedules (tiling, fusion, parallelization) on CPUs and GPUs.
However, these compilers target CPUs and GPUs---none currently generates code for tile-based AI accelerators.
%
%Ansor explicitly notes that it ``comes short of utilizing special instructions, such as Intel VNNI \cite{vnni}, NVIDIA Tensor Core \cite{markidis2018nvidia}, and ARM Dot''---precisely the gap that \pname{}'s ISA synthesis addresses.
%
\pname{} jointly explores algebraic transformations and schedule transformations (tiling, fusion) for tile-based accelerator programs, whereas schedule-based compilers only explore schedules for a fixed algorithm.

\textbf{Synthesis-based compilers.}
%Program synthesis has been used for instruction selection on vector architectures: 
Works such as Rake~\cite{ahmad2022vector}, Pitchfork \cite{root2023fast}, Hydride~\cite{kothari2024hydride} and Misaal \cite{noor2025misaal} use program synthesis to synthesize for x86 \cite{vnni}, ARM \cite{armintrinsics}, and Hexagon DSP \cite{codrescu2015architecture} vector ISA; Diospyros~\cite{vanhattum2021vectorization} targets Tensilica DSP's vector  ISA.
%and Swizzle Inventor~\cite{phothilimthana2019swizzle} synthesizes data swizzles for GPU kernels.
%
However, all of these works only target 1-D vector instructions. % and do not handle the complex 2-D tile instructions essential for modern AI accelerators. 
%
%A recent exception is 
Hardboiled~\cite{zhang2026pushing} uses equality saturation in Halide~\cite{ragan2013halide} to target 2-D Intel AMX \cite{vnni} and Nvidia Tensor Core \cite{nvidiatensorcores} ISA for signal and image processing; however, it only lowers to matrix multiply and accumulate operations.
Whereas prior works on guided tensor program lifting~\cite{brauckmann2026tensor, li2025guided} synthesize higher-level abstractions from legacy C/C++ code, without synthesizing complex tensor operations, \pname{} synthesizes 2-D tile instructions with hardware-constrained tile dimensions. %, and uses SMT over unbounded tensors to check for correctness.

\textbf{Tile-based accelerator languages.}
Vendors provide tile-based kernel languages for their accelerators: NKI~\cite{aws_neuron_nki_index} for Amazon's Trainium, Pallas~\cite{pallas} for Google's TPUs, and Triton~\cite{tillet2019triton} for GPUs.
Higher-level frameworks such as Hidet~\cite{ding2023hidet}, TileLang~\cite{wang2025tilelang}, and Helion~\cite{helion} compile from high-level tensor programs to GPU kernels.
All of these require programmers to have knowledge of hardware-specific details.
\pname{} takes a high-level NumPy-like program and automatically generates optimized code in NKI, without requiring programmers to have hardware expertise.

\textbf{Verified tensor transformations.}
Liu et al.~\cite{liu2022verified} present a Coq framework that verifies \textit{scheduling} rewrites (tiling, fusion, compute\_at) for Halide programs via machine-checked proofs.
TensorRight~\cite{arora2025tensorright} provides an SMT-based framework for verifying XLA \cite{openxla} rewrite rules over unbounded tensors. 
NiceToMeetYou~\cite{peng2026nicetomeetyou} uses synthesis with SMT-based verification to generate sound abstract transformers for LLVM \cite{lattner2004llvm} static analyses from MLIR \cite{lattner2021mlir} semantics. \pname{} uses SMT-verified algebraic transformations and ISA instruction selection for AI accelerators.
%
%NiceToMeetYou~\cite{peng2026nicetomeetyou} applies synthesis with SMT-based verification to a different compiler component, automatically generating sound abstract transformers for LLVM's static analyses from MLIR instruction semantics.
%
%\pname{} adapts a similar synthesis-plus-verification paradigm---using SMT to check correctness of synthesized results---but targets tensor program transformations and ISA instruction selection for AI accelerators.
\section{Conclusion}
\label{sec:conclusion}

%We introduced \pname{}, a synthesizing superoptimizer for tile-based AI accelerator programs that takes high-level tensor programs and automatically generates optimized kernels for the target hardware.
%
%\pname{} combines provably correct algebraic transformations via operator propagation, program synthesis for ISA instruction selection, symbolic tiling exploration, and operator fusion---maintaining all semantically equivalent program variants simultaneously in a $\nu$Graph and selecting the best empirically.
%
We introduced \pname{}, a synthesizing superoptimizer that automatically generates optimized tile-based AI accelerator kernels from high-level tensor programs, combining provably correct algebraic transformations, ISA synthesis, tiling, and fusion in a unified $\nu$Graph representation.
On Amazon's Trainium across 20 benchmarks, \pname{} achieves speedups up to 3.7x on individual operators and up to 19x on multi-operator kernels over the Neuron compiler, up to 1.35x over hand-optimized NKI kernels, and geomean speedups of 2--10\% over Mirage on all benchmarks it supports. %---while synthesizing 16 additional kernels that Mirage cannot.

%\section{Data-Availability Statement}
%The code is not available during the double-blind review period. If the paper is accepted, we will release the complete implementation as open-source.
%\chungha{Please write this down.}

\bibliographystyle{ACM-Reference-Format}
\bibliography{refs}
\newpage
\appendix

\section{Threats to Validity}
\label{app:threats}

\paragraph{End-to-end inference.}
Our evaluation measures kernel-level performance, consistent with Mirage~\cite{wu2025mirage} and other kernel optimization work.
End-to-end inference latency depends on additional factors such as kernel launch overhead, collective communication, and framework-level scheduling, which are orthogonal to the kernel optimizations presented in this paper.

\paragraph{Generalizability beyond Trainium.}
\pname{}'s core algorithms---algebraic transformations, tiling exploration, ISA synthesis, and fusion---could be adapted to other tile-based accelerators with similar architectural properties (Section~\ref{sec:background}).
Targeting a new platform requires providing new ISA semantics, hardware constraints, and a code emitter; the synthesis pipeline itself remains unchanged.
Accelerators with fundamentally different execution models (e.g., many-core architectures, GPU thread-block models) are not directly targeted by \pname{}'s current design.

\paragraph{Hardware generation.}
We evaluate on first-generation Trainium due to limited public availability of newer generations (Trainium2~\cite{aws_neuron_trainium2} and Trainium3~\cite{aws_neuron_trainium3}).
Based on published architecture descriptions, all Trainium generations share the same multi-engine design, NKI programming model, and tile-based execution model, so we expect \pname{}'s optimizations to transfer, though specific tiling parameters and performance characteristics may differ.

\paragraph{Benchmark coverage.}
Our benchmark suite covers core operators and multi-operator kernels, including LLM-specific layers such as GQA and Gated MLP, and overlaps significantly with Mirage's evaluation.
However, it does not cover all kernel types, such as convolutions, sparse operators, or training-specific kernels like backward passes.

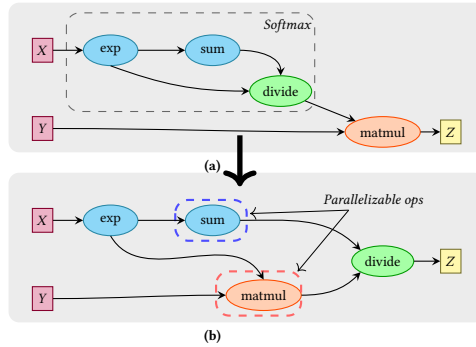
\begin{figure}[h]
\centering
\raisebox{0pt}[\height][0pt]{\begin{tikzpicture}[
    scale=0.45,
    transform shape,
    input/.style={rectangle, draw=purple!70!black, fill=purple!30, minimum width=0.6cm, minimum height=0.7cm, font=\Large},
    output/.style={rectangle, draw=olive!70!black, fill=yellow!40, minimum width=0.6cm, minimum height=0.7cm, font=\Large},
    softmaxop/.style={ellipse, draw=cyan!60!black, fill=cyan!40, minimum width=1.6cm, minimum height=0.9cm, font=\Large},
    mulop/.style={ellipse, draw=green!60!black, fill=green!35, minimum width=1.6cm, minimum height=0.9cm, font=\Large},
    matmulop/.style={ellipse, draw=red!40!orange, fill=red!25!orange!30, minimum width=1.6cm, minimum height=0.9cm, font=\Large},
    ghostop/.style={ellipse, draw=gray!60, dashed, fill=darkgray!20, minimum width=1.6cm, minimum height=0.9cm, font=\Large, text=black!70},
    arrow/.style={->, >=stealth},
    redarrow/.style={->, >=stealth, red!70, dashed},
    dashbox/.style={draw, dashed, rounded corners=5pt, inner sep=6pt}
]

% Part (a)
\begin{scope}[local bounding box=parta]
    \fill[gray!15, rounded corners] (1,1.4) rectangle (15,-3);

    %\node[input] (eps) at (0, 0.7) {$\epsilon$};
    \node[input] (x1) at (2, 0) {$X$};
    \node[input] (y1) at (2, -2.3) {$Y$};

    \node[softmaxop] (exp) at (4, 0) {exp};
    \node[softmaxop] (sum) at (7, 0) {sum};
    %\node[softmaxop] (div) at (6, 0) {div};
    %\node[softmaxop] (rsqrt) at (8, 0) {rsqrt};

    \node[mulop] (div) at (9, -1.2) {divide};
    \node[matmulop] (mm) at (12, -2.4) {matmul};
    \node[output] (z1) at (14, -2.4) {$Z$};

    \draw[dashbox, black!60] (2.7, 1.1) rectangle (10, -1.8);
    \node[font=\Large\itshape, anchor=north east] at (10, 1.05) {Softmax};

    \draw[arrow] (x1.east) to[out=0, in=180] (exp.west);
    \draw[arrow] (exp) -- (sum);
    %\draw[arrow] (sum) -- (div);
    %\draw[arrow] (eps.east) to[out=10, in=160] (add.north west);
    %\draw[arrow] (add) -- (rsqrt);
    \draw[arrow] (sum.east) to[out=0, in=90] (div.north);
    \draw[arrow] (exp.south) to[out=-20, in=180] (div.west);
    \draw[arrow] (div) -- (mm);
    \draw[arrow] (y1.east) to[out=0, in=180] (mm.west);
    \draw[arrow] (mm) -- (z1);

    \node[font=\Large\bfseries] at (7, -3.4) {(a)};
\end{scope}

% Part (b)
\begin{scope}[yshift=-5cm, local bounding box=partb]
    \fill[gray!15, rounded corners] (1,1.4) rectangle (15,-3);

    \node[font=\Large\itshape, align=center, anchor=north east] at (13.3, 0.9) {Parallelizable ops};

    %\node[input] (eps3) at (0, 0.7) {$\epsilon$};
    \node[input] (x3) at (2, 0) {$X$};
    \node[input] (y3) at (2, -2.3) {$Y$};

    \node[softmaxop] (exp3) at (4, 0) {exp};
    \node[softmaxop] (sum3) at (7, 0) {sum};
    %\node[softmaxop] (add3) at (6, 0) {add};
    %\node[softmaxop] (rsqrt3) at (8, 0) {rsqrt};

    \draw[dashbox, blue!70, thick] (5.9, 0.6) rectangle (8, -0.6);

    \node[matmulop] (mm3) at (8.5, -2.2) {matmul};
    \draw[dashbox, red!60, thick] (7.1, -1.5) rectangle (9.6, -2.8);

    \node[mulop] (div3) at (12, -1.2) {divide};
    \node[output] (z3) at (14, -1.2) {$Z$};

   \draw[->, line width=2pt] (7.8,2.5) -- (7.8,1);

   \draw[->, line width=0.3pt] (11,0.3) -- (9.5, -1.5);

   \draw[->, line width=0.3pt] (11,0.3) -- (8.1, 0.2);

    \draw[arrow] (x3.east) to[out=0, in=180] (exp3.west);
    \draw[arrow] (exp3) -- (sum3);
    %\draw[arrow] (mean3) -- (add3);
    %\draw[arrow] (eps3.east) to[out=10, in=160] (add3.north west);
    %\draw[arrow] (add3) -- (rsqrt3);
    \draw[arrow] (sum3.east) to[out=0, in=135] (div3.north west);
    \draw[arrow] (exp3.south) to[out=-60, in=120] (mm3.north);
    \draw[arrow] (y3.east) to[out=0, in=180] (mm3.west);
    \draw[arrow] (mm3.east) to[out=0, in=-135] (div3.south west);
    \draw[arrow] (div3) -- (z3);

    \node[font=\Large\bfseries] at (7, -3.4) {(b)};
\end{scope}
% Clip bounding box to remove extra whitespace
\pgfresetboundingbox
\useasboundingbox (1,1.4) rectangle (15,-7.8);
\end{tikzpicture}}
\caption{(a) Computation graph for Softmax+MatMul. (b) Semantically equivalent Softmax+MatMul graph with two parallelizable operations.}
\label{fig:softmax-matmul-comp-graph}
\end{figure}

\section{Additional Algebraic Transformation Examples}
\label{app:additional-examples}

\subsection{Softmax+MatMul}
\label{app:softmax-matmul}

Figure~\ref{fig:softmax-matmul-comp-graph} shows a Softmax+MatMul kernel where the \texttt{divide} and \texttt{matmul} operators can be reordered so that \texttt{matmul} executes on the Tensor Engine while the row-wise \texttt{sum} reduction executes on the Vector Engine in parallel, rather than sequentially.

This transformation is analogous to the RMSNorm+MatMul rewrite discussed in Section~\ref{sec:motivation}: an element-wise operation (here, \texttt{divide} by the softmax normalization factor) commutes with \texttt{matmul} because the normalization factor reduces along the same dimension contracted by the matrix multiplication.

\pname{} automatically discovers this transformation through operator propagation and provably checks it using SMT over unbounded tensors (Section~\ref{sec:verification}).
However, existing graph rewriting systems---including TASO~\cite{jia2019taso}, Tensat~\cite{tensat}, Constable~\cite{vohra2025mind}, and Mirage~\cite{wu2025mirage}---do not support this rewrite because their hand-written rules do not cover this particular pattern of operator reordering.
This illustrates the fundamental limitation of pattern-matching approaches: they can only perform transformations for patterns that have been manually anticipated and implemented.

\section{Ablation Study: Per-Dimension Results}
\label{app:ablation-figures}

\begin{figure*}[h]
  \centering
  \begin{minipage}[b]{0.99\textwidth}
    \centering
    \includegraphics[width=0.85\linewidth]{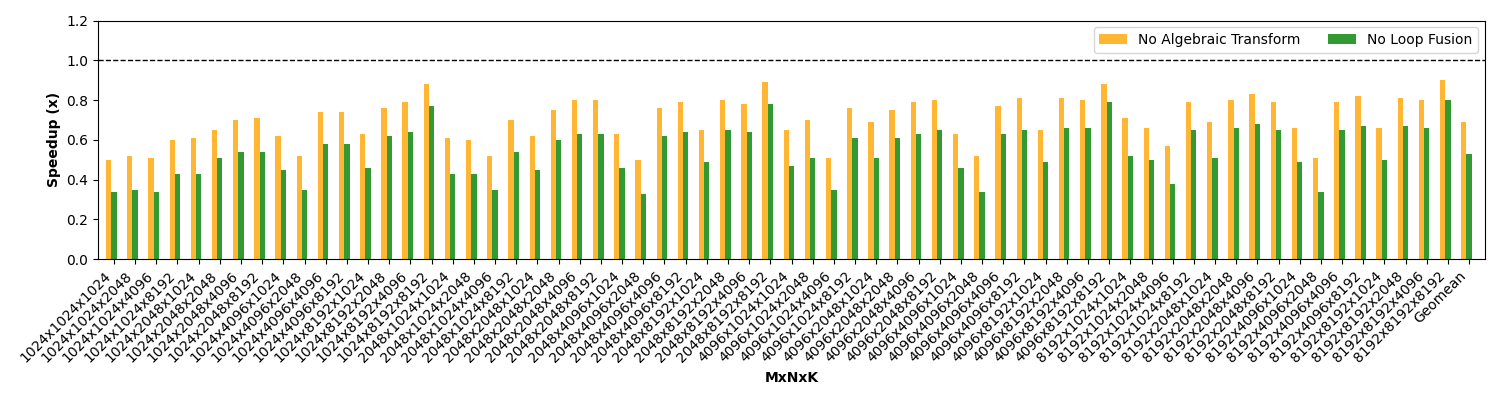}
    \vspace{-5pt}
    \subcaption{RMSNorm + MatMul}
  \end{minipage}
  \vspace{-3pt}
  \begin{minipage}[b]{0.99\textwidth}
    \centering
    \includegraphics[width=0.85\linewidth]{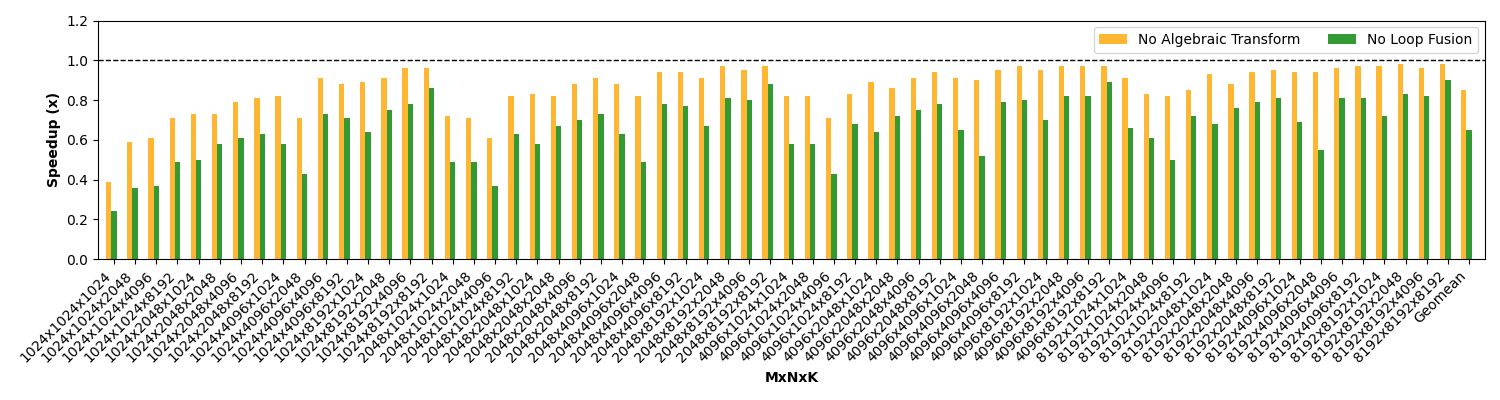}
    \vspace{-5pt}
    \subcaption{QKV Projection}
  \end{minipage}
  \vspace{-3pt}
  \begin{minipage}[b]{0.99\textwidth}
    \centering
    \includegraphics[width=0.85\linewidth]{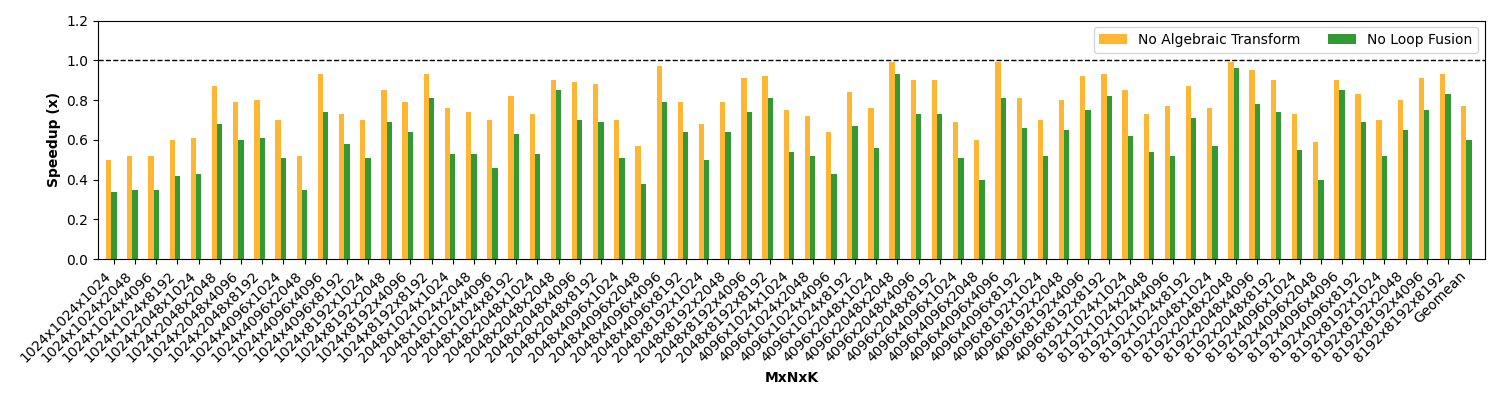}
    \vspace{-5pt}
    \subcaption{Group Query Attention}
  \end{minipage}
  \vspace{-3pt}
  \begin{minipage}[b]{0.99\textwidth}
    \centering
    \includegraphics[width=0.85\linewidth]{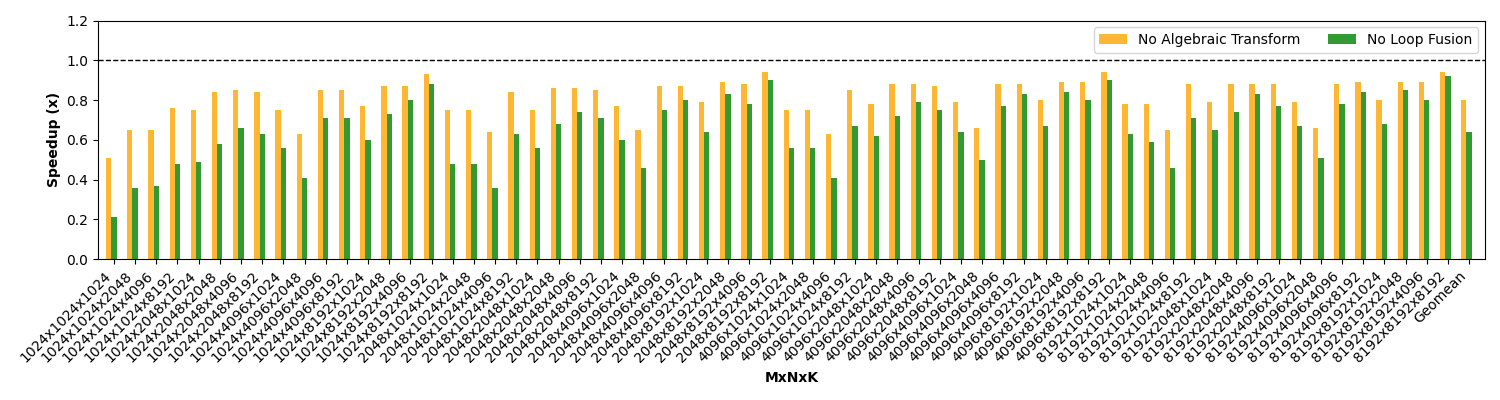}
    \vspace{-5pt}
    \subcaption{Gated MLP}
  \end{minipage}
  \caption{Impact of disabling algebraic transformations and fusion on four multi-operator benchmarks across all dimension sizes from Table~\ref{tab:multi-op-eval}.}
  \label{fig:ablation-detail}
\end{figure*}

Figure~\ref{fig:ablation-detail} shows the per-dimension results for the ablation study discussed in Section~\ref{subsec:ablation}. Each bar shows performance relative to \pname{} with all optimizations enabled (dashed line at 1.0x).

\section{Generated NKI Kernel for RMSNorm+MatMul}
\label{app:generated-nki}

Figure~\ref{fig:generated-nki} shows the NKI kernel generated by \pname{} for the RMSNorm+MatMul input program from Figure~\ref{fig:ap-example}. The RMS branch (lines 18--21) executes on the Vector/Scalar Engine while the MatMul branch (lines 24--39) executes on the Tensor Engine in parallel within the same loop nest. The results are combined at the end (lines 46--49), reflecting the algebraic transformation from Figure~\ref{fig:rmsnorm-matmul-comp-graph}(b).

\begin{figure}[h]
\centering
\scalebox{0.99}{
\input{figures/generated_nki.tex}
}
\caption{NKI kernel generated by \pname{} for RMSNorm+MatMul from the 5-line input in Figure~\ref{fig:ap-example}.}
\label{fig:generated-nki}
\end{figure}

\begin{comment}

\begin{figure*}[t]
\centering
\vspace{-20pt}
\scalebox{0.8}{
\raisebox{0pt}[\height][0pt]{\input{figures/fused_example.tex}}
}
\caption{An example of fusion of \texttt{nc\_transpose} and \texttt{transpose} operation when emitting NKI code.}
%\vspace{-20pt}
\label{fig:fused-isa-graph}
\end{figure*}
\end{comment}

\end{document}